\documentclass[twocolumn,prc,showpacs,preprintnumbers] {revtex4-2}
\usepackage{graphicx,latexsym,amssymb,amsmath,color,multirow,mathrsfs,booktabs,ifpdf,epsfig,dcolumn,bm,longtable}
\usepackage{changes}
\usepackage{hyperref}
\hypersetup{colorlinks=true,linkcolor=blue,filecolor=magenta,urlcolor=cyan}
\date{\today}

\begin{document}
\title{Potential model with bound-to-continuum approach for low-energy nucleon radiative capture by $^{12}$C and $^{16}$O}
    \renewcommand{\andname}{\ignorespaces}
	\author{Nguyen Le Anh$^{1,2,3}$}
	\email{anhnl@hcmue.edu.vn}
	\affiliation{$^{1}$Department of Theoretical Physics, Faculty of Physics and Engineering Physics, University of Science, Ho Chi Minh City, Vietnam.}
	\affiliation{$^{2}$Vietnam National University, Ho Chi Minh City, Vietnam.}
	\affiliation{$^{3}$Department of Physics, Ho Chi Minh City University of Education, 280 An Duong Vuong, District 5, Ho Chi Minh City, Vietnam.}
	\author{Phan Nhut Huan$^{4,5}$}
	\email{phannhuthuan@duytan.edu.vn}
	\affiliation{$^{4}$Institute of Fundamental and Applied Sciences, Duy Tan University, Ho Chi Minh City, Vietnam.}
    \affiliation{$^{5}$Faculty of Natural Sciences, Duy Tan University, Da Nang City, Vietnam.}
	\author{Bui Minh Loc$^{6,7}$}
	\email{Corresponding author: buiminhloc@tdtu.edu.vn}
	\affiliation{$^{6}$Division of Nuclear Physics, Advanced Institute of Materials Science, Ton Duc Thang University, Ho Chi Minh City, Vietnam.}
    \affiliation{$^{7}$Faculty of Applied Sciences, Ton Duc Thang University, Ho Chi Minh City, Vietnam.}

\begin{abstract}
The nucleon radiative capture reactions are important in pure and applied nuclear physics, especially in nuclear astrophysics.
The keV-nucleon radiative capture reactions are studied with $^{12}$C and $^{16}$O targets using the bound-to-continuum potential model in which both scattering and bound states are treated simultaneously and based on the Skyrme Hartree-Fock approximation. 
The obtained results are shown to be in good agreement with the available experimental data. Alongside astrophysical aspects, the nuclear structure features were revisited for enlarging the prospect of adopting the nucleon radiative capture processes as a spectroscopic tool. 
\end{abstract}

\maketitle

\section{INTRODUCTION} \label{sec1}
The radiative capture (RC) reactions in which the incident particle such as neutron, proton, alpha, and other light ions is absorbed by the target nucleus, and the gamma radiation is then detected are important in applied and pure nuclear physics \cite{rol90}. In particular, the nucleon RC reaction is one of the most important in the formation of various elements in the universe, it is therefore essential in nuclear astrophysics \cite{bru15, nun20, des20}.
For example, the proton RC or $(p,\gamma)$ reactions appear in the CNO cycle, where they generate nuclear energy in massive stars \cite{bur57,arc17} and as a source of solar neutrinos \cite{vil21}. While the neutron capture reactions $(n,\gamma)$ play a key role in the nucleosynthesis on the inhomogeneous Big Bang model \cite{obe96,kaj02} and in the $s$-process in asymptotic giant branch (AGB) stars \cite{bus99}. Among different nuclei in the processes, $^{12}$C and $^{16}$O isotopes are particularly interested.

For the ($p,\gamma$) reaction, the $^{16}$O($p,\gamma$)$^{17}$F is the slowest process in all proton RC reactions in the CNO cycle because of the absence of low-lying resonances. The $^{12}$C($p,\gamma$)$^{13}$N reaction displays the next-to-slowest reaction rate in the CNO cycle. This reaction affects the rate in the outer parts of the solar core, where the CNO cycle has not yet reached its equilibrium at the lower temperature. 
For the ($n,\gamma$) reaction, the calculated rates implied from cross-sections of both $^{12}$C($n,\gamma$)$^{13}$C and $^{16}$O($n,\gamma$)$^{17}$O reactions are important concerning the nucleosynthesis in inhomogeneous Big Bang model and the $s$-process. Although the nucleus $^{16}$O has a small neutron-capture cross-section, the $^{16}$O plays an important role as a neutron poison in the astrophysical $s$-process due to its high abundance \cite{moh16}.

It is difficult to measure the cross-sections for nucleon RC reactions at low energy because of their extremely small reaction cross-section. Theoretical insight is therefore required to reach relevant stellar energies. Extrapolation of experimental RC reaction cross-sections downward from accessible to astrophysical energies relies on theoretical models \cite{car17,des20}.
Theoretically, the RC process is considered as electromagnetic transitions from the scattering states to the bound states. Among different theoretical models in the study of RC reactions (see Ref.~\cite{des20} for a review), the potential model is the simplest but powerful tool. It is commonly approached phenomenologically with Gaussian or Woods-Saxon potentials to obtain the nuclear wave functions \cite{ho83,ho91,bay04,hua10,dub13,xu13}. The microscopic nucleon-nucleus potential within the folding model was also applied \cite{kit02, anh21NPA} to reduce the number of free parameters that lead to the improvement in the prediction of cross-sections and reaction rates of the RC reactions. However, the folding model is suitable for the scattering problem only, and its application at very low energies contains impediments.

The RC process is an example where the scattering and bound wave functions of the same nucleus work together. However, they are often treated separately. Moreover, in previous studies of RC reactions, such as Refs.~\cite{hua10, dub13, anh21NPA}, the single-particle bound states were obtained with a phenomenological or folding model potential that does not relate to any nuclear structure model. To obtain the nuclear wave functions consistently and microscopically, the alpha-cluster description can be applied \cite{duf97}. This approach is suitable for the nuclear state described by the alpha-cluster model. The work in Ref.~\cite{anh21PRC} attempted to use a single mean-field for scattering and bound state within the potential model.

In the present work, the bound-to-continuum approach in Ref.~\cite{anh21PRC} was applied to obtain simultaneously the bound states and the scattering states. The approach is based on the single Skyrme Hartree-Fock (HF) calculation using the Skyrme interaction SLy4 \cite{cha98,cha03} with the corrected asymptotic forms of not only the scattering but also the bound state. The advantage of this approach is that there are only two well-constrained parameters with one purpose that is ensuring the asymptotic behavior of the wave functions. The approach can be applied simultaneously for the ($p,\gamma$) and ($n,\gamma$) reactions. The information from the ($p,\gamma$) reaction can be applied to the ($n,\gamma$) reactions that are usually poor in experimental data. In reverse, the neutron-induced reactions in general and ($n,\gamma$) reaction, in particular, provides additional information for the nuclear structure. The use of RC reactions as a tool for nuclear spectroscopy was discussed in great details in Ref.~\cite{rol73}. Therefore, not only were the experimental data well-reproduced but also the nuclear structure information is revealed in the bound-to-continuum approach.

\section{Formalism and method} \label{sec2}
\subsection{Potential model for nucleon radiative capture}
In the study of the RC reactions at low energy, it is customary to calculate the energy-dependent astrophysical $\mathcal{S}(E)$ factor as
\begin{equation} \label{sfactor}
    \mathcal{S}(E) = E \exp(2\pi\eta) \sigma(E),
\end{equation}
where $E$ is the energy of the incident nucleon, and $\eta$ is the Sommerfeld parameter. The capture cross-section for emission of electric dipole radiation ($E1$) in the transition $i \to f$ of the target which has the charge number $Z$ and the mass number $A$ is given by
\begin{equation} \label{xsection}
\sigma (E) = \frac{4}{3}\frac{e^2}{\hbar} \left(\dfrac{4\pi}{3}k^3_\gamma\right) \left(\dfrac{ A\tau - Z}{A + 1}\right)^2|\mathcal{M}^{E1}_{i \to f}|^2, 
\end{equation}
in which $\tau = 0$ for neutron, and $\tau = 1$ for proton. The emitted $\gamma$-ray wave number is
\begin{equation}
    k_{\gamma} = [E - (-Q + E_x)]/(\hbar c),
\end{equation}
where $E_x$ is the excitation energy of the daughter nucleus, and $Q$ is the $Q$-value of the reaction. Only the electric dipole $E1$ transition is considered in the study.

The matrix element $\mathcal{M}^{E1}_{i\to f}$ in Eq.~\eqref{xsection} can be decomposed into three components
\begin{equation}\label{3factors}
 \mathcal{M}^{E1}_{i\to f} = \mathcal{A} \cdot \mathcal{I} \cdot S_F.
\end{equation}
The first component $\mathcal{A}$ is the angular-spin coefficient given by \cite{men95}
\begin{align} 
\mathcal{A}^2_{i \to f} =\, \dfrac{3}{4\pi}(2\ell_i +1)(2j_i+1)(2j_f+1)(2J_f+1) \nonumber \\
\times (\ell_i 0 1 0|\ell_f 0)^2 \left\{ \begin{matrix}
j_i&I&J_i\\
J_f &1&j_f
\end{matrix} \right\}^2 \left\{ \begin{matrix}
\ell_i &1/2 &j_i\\
j_f &1 & \ell_f
\end{matrix} \right\}^2, \label{ang}
\end{align}
where the last round bracket component is a Clebsch-Gordan coefficient, and the last two curly bracket components are Wigner $6j$ coefficients. The nucleon spin and angular momentum are $\bm{s}$ and $\bm{\ell}$, respectively, and the total angular momentum is $\bm{j} = \bm{\ell} + \bm{s}$. The target is the inert core after the capture process with the unchanged spin $\bm{I}$. The total spin of the system is $\bm{J} = \bm{j} + \bm{I}$.
The second component that is the central part of the potential model is the radial overlap of the scattering state $\chi_{\ell_i}$ and bound state $\varphi_{n_f \ell_f j_f}$
\begin{equation} \label{overlap}
\mathcal{I}(E) = \int \varphi_{n_f \ell_f j_f}(r) \chi_{\ell_i} (E,r) r\,dr.
\end{equation}
The final component $S_F$ is the spectroscopic factor of $\varphi_{n_f \ell_f j_f}$.

\subsection{The bound-to-continuum potential model}
In the bound-to-continuum potential model \cite{anh21PRC}, both wave functions $\varphi_{n_f \ell_f j_f}(r)$ and $\chi_{\ell_i} (E,r)$ in Eq.~\eqref{overlap} were calculated from the Skyrme HF approximation. The calculation was started with the radial HF equations following Refs.~\cite{dov71, dov72},
\begin{eqnarray} \label{HFeq}
&& \left\{\dfrac{\hbar^2}{2m^*_\tau(r)}\left[- \frac{d^2}{dr^2} + \dfrac{\ell(\ell + 1)}{r^2} \right] + V^\tau(r) \nonumber \right.\\ && \left. - \frac{d}{dr}\left[\frac{\hbar^2}{2m^*_\tau(r)}\right] \frac{d }{dr} \right\}\varphi_\alpha(r) = \epsilon_\alpha \varphi_\alpha(r),
\end{eqnarray}
in which $\varphi_\alpha(r)$ and $\epsilon_\alpha$ are the HF single-particle wave function and energy with $\alpha$ being the set of all necessary quantum numbers, respectively. $V^\tau(r)$ is the Skyrme HF potential \cite{dov71, vau72} including the central $V^\tau_{\rm c}(r)$, the spin-orbit $V^\tau_{\rm s.o.}(r)$, and the one-body Coulomb potential $\tau V_{\rm Coul.}(r)$. The first-order derivative term in Eq.~\eqref{HFeq} is eliminated by using the transformation
\begin{equation}\label{Dovertrans}
    \varphi_\alpha(r) = [m^*_\tau(r)/m]^{1/2}\tilde{\varphi}_\alpha(r),
\end{equation}
then Eq.~\eqref{HFeq} can be rewritten in terms of the usual Schr\"odinger equation. Ref.~\cite{col13} presented a tool for the Skyrme HF calculation in detail that is used in our study. After the iteration in calculation, the converged potentials $V^\tau(r)$ and effective masses $m^*_\tau(r)$ are used to determine the occupied and unoccupied single-particle bound states. In the study, the unoccupied single-particle bound states of interest are given in Table \ref{tab1}.

Following Refs.~\cite{dov71, dov72}, Eq.~\eqref{HFeq} is also applied for the scattering state with the discrete eigenvalue $\epsilon_\alpha$ replaced by the continuous variable $E$.
Consequently, the bound and scattering wave functions in the overlap integral in Eq.~\eqref{overlap} are solutions of
\begin{align}
\left\{\dfrac{\hbar^2}{2 m'}\left[-\dfrac{d^2}{dr^2} + \dfrac{\ell_f(\ell_f + 1)}{r^2} \right] + \mathcal{V}_b(\epsilon_\alpha, r) -\epsilon_\alpha \right\} \nonumber \\ \tilde{\varphi}_\alpha(r) =  0, \label{SEb}
\\
\left\{\dfrac{\hbar^2}{2 m'} \left[-\dfrac{d^2}{dr^2} + \dfrac{\ell_i(\ell_i + 1)}{r^2} \right] +\mathcal{V}_s(E, r) -E \right\} \nonumber \\ \chi_{l_i}(E, r) = 0, \label{SEs}
\end{align}
where now $\alpha \equiv n_f \ell_f j_f$, for example $1d_{5/2}$ in the case of $^{16}$O($p,\gamma$)$^{17}$F reaction. In both Eqs.~\eqref{SEb} and \eqref{SEs}, $m' = m A/(A-1)$ with $m$ being the nucleon mass. The use of $m'$ is to take into account a large part of the center-of-mass (c.m.) correction. It is important for light nuclei in the study. Note that $\tilde{\varphi}_\alpha(r)$ is transformed back to $\varphi_\alpha(r)$ by Eq.~\eqref{Dovertrans}, while the scattering wave function $\chi_l(E, r)$ is kept. The potentials $\mathcal{V}_b(\epsilon_\alpha,r)$ and $\mathcal{V}_s(E,r)$ in Eqs. \eqref{SEb} and \eqref{SEs} are
\begin{align}
    \mathcal{V}_b (\epsilon_\alpha,r) &= \lambda_b \mathcal{V}_b^{\rm c} (\epsilon_\alpha,r) + \mathcal{V}_{\rm Coul.}(r) + \mathcal{V}_{\rm s.o.}(r)\bm{\ell}\cdot\bm{\sigma} , \label{lambdab}\\
    \mathcal{V}_s (E,r) &= \lambda_s \mathcal{V}_s^{\rm c} (E,r) + \mathcal{V}_{\rm Coul.}(r) + \mathcal{V}_{\rm s.o.}(r)\bm{\ell}\cdot\bm{\sigma}. \label{lambdas}
\end{align}
$\mathcal{V}^{\rm c}_b(\epsilon_\alpha, r)$ and $\mathcal{V}^{\rm c}_s(E, r)$ are energy-dependent central potentials, particularly
\begin{eqnarray} \label{pot}
\mathcal{V}^{\rm c}_b(\epsilon_\alpha, r) = \dfrac{m^*_\tau(r)}{m} \Bigg\{V_{\rm c}^{\tau}(r) + \dfrac{1}{2} \left(\dfrac{\hbar^2}{2m^*_\tau(r)} \right)'' \nonumber \\ 
- \dfrac{m^*_\tau(r)}{2\hbar^2} \left[\left(\dfrac{\hbar^2}{2m^*_\tau(r)}\right)'\right]^2\Bigg\} + \left[1 - \dfrac{m^*_\tau(r)}{m}\right] \epsilon_\alpha, 
\end{eqnarray}
and $\mathcal{V}^{\rm c}_s(E, r)$ is obtained using the continuous incident energy $E$ instead of the single-particle energy $\epsilon_\alpha$. The spin-orbit and Coulomb potentials are included
\begin{eqnarray}
    \mathcal{V}_{\rm Coul.}(r) &=& [m^*_{\tau =1}(r)/m] V_{\rm Coul.}(r), \label{Vcoul}\\ 
    \mathcal{V}_{\rm s.o.}(r) &=& [m^*_{\tau}(r)/m] V^\tau_{\rm s.o.}(r).
\end{eqnarray}
Unlike the central potential, they do not depend on the energy, hence the same in Eq.~\eqref{SEb} and Eq.~\eqref{SEs}. Note that only the depths of the nuclear central parts are modified. The spin-orbit and Coulomb potentials are kept unchanged. The two parameters in the study are $\lambda_b$ and $\lambda_s$ in Eqs.~\eqref{lambdab} and \eqref{lambdas}, respectively.

For the scattering state, the parameter $\lambda_s$ is introduced to be adjusted such that the position of a low-lying resonance is reproduced \cite{anh21PRC}. The asymptotic behavior of the scattering wave function is therefore verified.
For the bound state, the correct asymptotic behavior of the unoccupied bound wave functions $\varphi_{n_f \ell_f j_f}(r)$ is also essential. In our approach, the well-depth method is applied for generating the correct asymptotic form of the single-particle bound state $\varphi_{n_f \ell_f j_f}(r)$ obtained in Eq.~\eqref{SEb}. The nuclear central potential is multiplied by the scaling factor $\lambda_b$ such that $\epsilon_{n_f \ell_f j_f}$ equals to the value $e_s$ that is in most cases the experimental nucleon separation energy $S_{p(n)} = -Q + E_x$. The $Q$-values and excitation energies $E_x$ are taken from the experimental nuclear database \cite{sel91,til93,wan12}. The values of $e_s$ and $S_{p(n)}$ are presented in Table \ref{tab1} for each state. Note that the occupied single-particle states are kept unchanged as the inert core.

Finally, the spectroscopic factor $S_F$ is varied to reproduce the experimental data. However, it is not considered as a parameter because its value is traditionally constrained by the $(d,p)$ stripping reactions \cite{mac60} or the shell-model calculations \cite{coh67}. For the nuclei in the study, $S_F$ should be close to unity for the case of single-particle states in the $sd$-space, and it is far from unity in the case of single-particle states in the $p$-space.

Consequently, $V^\tau(r)$ and $m_\tau^*(r)$ in the Skyrme HF approximation can be used to obtain simultaneously the bound state and the scattering state in the study of keV-nucleon RC reaction with only two parameters. The two parameters simply adjust the depth of the central nuclear potentials to verify the asymptotic behavior of the wave functions. The scaling factor $\lambda_b$ for the bound state is constrained by the nucleon separation energy, and $\lambda_s$ for the scattering state is determined by the position of the low-lying resonance.

\section{Results and discussions} \label{sec3}
First, the ($p,\gamma$) and ($n,\gamma$) on the double-closed-shell nucleus $^{16}$O are discussed. Then with the adjustments of $\lambda_s$, $\lambda_b$, and $S_F$, the method is applied for the case of $^{12}$C. The summary of all calculations is presented in Table \ref{tab1}.
\begin{table}[t]
    \centering
        \caption{The main transitions from the scattering state $\chi_{l_{i}}$ to the bound state $\varphi_\alpha$ with $\alpha \equiv \{n_f \ell_f j_f\}$, and $\epsilon_\alpha$ (in MeV) being its single-particle energy. $J_f^\pi$ is the total spin of the final state. $e_s$ (in MeV) is the nucleon separation energy used in our calculation, while $S_{p(n)} = -Q + E_x$ (in MeV) is the experimental proton (neutron) separation energy \cite{sel91,til93,wan12}. $\lambda_b$ is the scaling factor of the bound potential. $S_F$ is the spectroscopic factor.}
    \begin{tabular}{llllrrrrr}
    \hline \hline
	Reaction & $J_f^\pi$	&	$\chi_{l_{i}}$	&	$\varphi_\alpha$	&	$\epsilon_\alpha$ & $S_{p(n)}$	& $e_s$	&	$\lambda_b$	&	$S_F$	\\
\hline
	$^{16}$O($p,\gamma$)$^{17}$F & $\frac{5}{2}^+$	&	$p,f$	&	1$d_{\frac{5}{2}}$	&	$-3.57$	&  $-0.60$ &	$-0.60$	&	$0.88$	&	$1.00$	\\
	$^{16}$O($p,\gamma$)$^{17}$F$^{*}$ & $\frac{1}{2}^+$	&	$p$	&	2$s_{\frac{1}{2}}$	&	$-0.97$	& $-0.11$	& $-0.11$	&	$0.95$	&	$1.00$	\\
	$^{16}$O($n,\gamma$)$^{17}$O & $\frac{5}{2}^+$	&	$p,f$	&	1$d_{\frac{5}{2}}$	&	$-6.75$	& $-4.14$ &	$-4.14$	&	$0.90$	&	$1.00$	\\
	$^{16}$O($n,\gamma$)$^{17}$O$^{*}$ & $\frac{1}{2}^+$		&	$p$	&	2$s_{\frac{1}{2}}$	&	$-3.89$	& $-3.27$	&	$-3.89$	&	$1.00$	&	$1.00$	\\
	$^{12}$C($p,\gamma$)$^{13}$N &$\frac{1}{2}^-$	&	$s,d$	&	1$p_{\frac{1}{2}}$	&	$-6.85$	& $-1.94$	&	$-1.94$	&	$0.81$	&	$0.20$	\\
	$^{12}$C($n,\gamma$)$^{13}$C & $\frac{1}{2}^-$	&	$s,d$	&	1$p_{\frac{1}{2}}$	&	$-9.42$	& $-4.95$	&	$-4.95$	&	$0.83$	&	$0.45$	\\
	$^{12}$C($n,\gamma$)$^{13}$C$^{*}$ & $\frac{1}{2}^+$	&	$p$	&	2$s_{\frac{1}{2}}$	&	$-1.25$&	$-1.86$	&	$-3.02$	&	$1.11$	&	$1.00$	\\
	$^{12}$C($n,\gamma$)$^{13}$C$^{*}$ & $\frac{3}{2}^-$	&	$s,d$	&	1$p_{\frac{3}{2}}$	&	$-16.85$&	$-1.26$	&	$-1.26$	&	$0.45$	&	$0.25$	\\
	  & 	&	$p,f$	&	1$d_{\frac{3}{2}}$	&	$0.97$	&	$-1.26$&	$-1.26$	&	$1.30$	&	$0.95$	\\
	$^{12}$C($n,\gamma$)$^{13}$C$^{*}$ & $\frac{5}{2}^+$	&	$p,f$	&	$1d_{\frac{5}{2}}$	&	$-2.30$	& $-1.09$	&	$-1.09$	&	$0.94$	&	$0.55$	\\
\hline \hline
    \end{tabular}
    \label{tab1}
\end{table}

The scattering state is well-described in our approach. The asymptotic form of the scattering state was verified by the studies of nucleon elastic scattering \cite{dov71, dov72, hao15}. The parameter $\lambda_s$ is adjusted such that the position of a low-lying resonance is reproduced. In the case of $^{12}$C($p, \gamma$), it is the resonance at $0.42$ MeV corresponding to the $1/2^+_1$ excited state at $2.365$ MeV in $^{13}$N structure. The value of $\lambda_s$ obtained from $^{12}$C($p, \gamma$) is then applied for $^{12}$C($n, \gamma$) reactions as they are non-resonant RC reactions. For the case of $^{16}$O, the lack of low-lying resonance is covered by the HF approximation that is well-applied for double-closed shell nuclei. The parameters $\lambda_s$ of the scattering state are equal to 1.00 and 1.02 in the cases of $^{16}$O and $^{12}$C, respectively. It is worth reminding some properties of the potential $\mathcal{V}_s(E, r)$ in our approach. It is purely real. The dispersive optical potential used in Ref.~\cite{kit98} for the $(n,\gamma)$ reaction was not included in our calculation. As $\lambda_s$ is close to unity, the dispersive component is less than 5\% and can be taken into account by the parameter $\lambda_s$. The non-locality of the optical potential in the RC reaction was mentioned in Ref.~\cite{tia18}. In the Skyrme-HF approximation, the non-locality of the mean-field was taken into account by the damping factor $m^*_\tau(r)/m$ and the energy-dependent term $[1 - m^*_\tau(r)/m]E$ \cite{dov72}. Note that the Coulomb potential is important to the low-energy scattering. It is different from the usual Coulomb potential because of the damping factor $m^*_{\tau =1}(r)/m$ in Eq.~\eqref{Vcoul}. The difference does not behave like $1/r$ for large distance, but has the shape of a Fermi function \cite{dov72}. Therefore, the difference does not affect the calculation for large distance.

The bound state is based on the Skyrme HF calculation including the correct asymptotic form. The value of $\lambda_b$ depends on the difference between the single-particle energy $\epsilon_\alpha$ calculated with $\lambda_b = 1.00$ and the nucleon separation energy, $\Delta e = e_s - \epsilon_\alpha$ (in MeV). From our calculations (Table \ref{tab1}), there is a linear correlation that is 
$(1 - \lambda_b) = 0.035\Delta e/\text{MeV}$. 
This formula is valid for all cases in the study, i.e. for $^{16}$O and $^{12}$C, and for the transitions to the ground state and to the excited state. It comes from the fact that the energy dependence of the potential produced by the Skyrme HF formalism is linear in energy in Eq.~\eqref{pot}.

The spectroscopic factors $S_F$ in our analysis are empirical values as it is varied to reproduce the experimental data. It contains partially the core-polarization effect induced by the incident particle that is neglected in the calculation. This effect is known to be small in nucleon elastic scattering. However, in the RC reaction, the situation is quite different. Indeed, the incident nucleon is just captured onto the bound state and it is waiting there until it will radiate the photon. The electromagnetic processes are much slower than those due to the strong interactions. Then it would mean that there is enough time for the captured nucleon to interact with the main part of the nucleus so that its wave function is affected by the core. Probably, this process is partially included in the empirical spectroscopic factors in our calculations. The process is less important for $^{16}$O, but it is certain for $^{12}$C. $S_F$ of $^{16}$O($n,\gamma$) reactions are unity. $S_F$ in the case of $^{12}$C($p, \gamma$) reactions are smaller than other calculations because of this effect in the RC reaction. With alpha-substructures, the effect in both nuclei is actually important, it is, however, not in the context of the nucleon mean-field approach in the study. Moreover, the proton interacts with the core rather than the neutron does because of the Coulomb interaction. That is the reason why the $S_F$ of ($p,\gamma$) reactions are smaller than those of ($n,\gamma$) reactions (see Table \ref{tab1}).

\subsection{$^{16}$O($p,\gamma$)$^{17}$F and $^{16}$O($n,\gamma$)$^{17}$O}
The result for $^{16}$O($p,\gamma$)$^{17}$F reaction was reported in Ref.~\cite{anh21PRC}, however, $^{16}$O($p,\gamma$)$^{17}$F and $^{16}$O($n,\gamma$)$^{17}$O were simultaneously analyzed in the present study. Both reactions are simple examples of the RC in nuclear astrophysics, although they are non-resonant reactions in the energy region of interest. The $^{16}$O isotope is tightly bound with the doubly-closed-shell structure that is well described in the HF approximation. The Skyrme HF for the continuum was successful for the nucleon elastic scattering at low energy, especially for $^{16}$O \cite{dov71}. The implication of this model is therefore expected for the study of both proton and neutron RC reactions at low energies for $^{16}$O. It is as one can expect that the spectroscopic factors $S_F$ of $1d_{5/2}$ and $2s_{1/2}$ are unity (Table \ref{tab1}). It means the nucleon wave functions are not affected by the core during the capture process in the case of $^{16}$O.

\begin{figure}[t]
    \includegraphics[width=1.\linewidth]{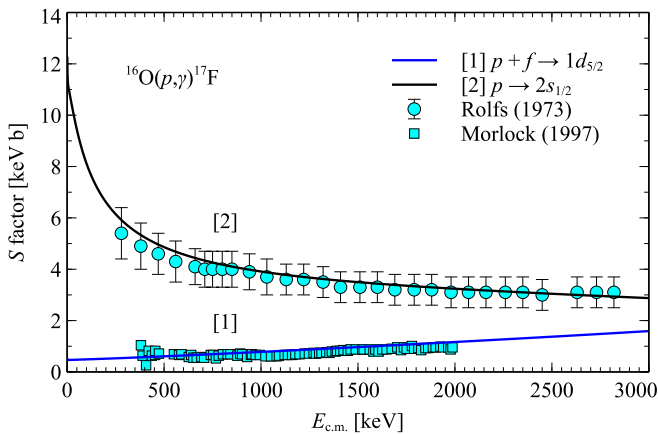}
    \caption{Calculated astrophysical $\mathcal{S}(E)$ of the $^{16}$O($p,\gamma$)$^{17}$F reaction. The transition to the ground state is noted as [1], and to the excited state [2]. The spectroscopic factors of $1d_{5/2}$ and $2s_{1/2}$ are $1.00$. Experimental data were taken from Ref.~\cite{mor97} for the transition to the ground state, and Ref.~\cite{rol73} for the transition to the excited state.}
    \label{16Opg}
\end{figure}

Indeed, in Fig.~\ref{16Opg} the calculated $\mathcal{S}(E)$ for the $^{16}$O($p,\gamma$)$^{17}$F reaction well reproduces the experimental data up to 2--3 MeV taken from Refs.~\cite{rol73,mor97}. The transitions to the ground state $^{17}$F (noted as [1] in Fig.~\ref{16Opg}) is dominated by $E1$ transitions from $p+f$-scattering states to the $1d_{5/2}$-bound state (Table \ref{tab1}). As the doubly-closed-shell structure of $^{16}$O, the structure of ground state $^{17}$F ($5/2^+$) is implied as the inert $^{16}$O-core and one additional proton in $1d_{5/2}$ state. Similarly, for the transition to the excited state $^{17}$F$^{*}$ ($1/2^+$) (noted as [2] in Fig.~\ref{16Opg}), the incident proton is captured to the $2s_{1/2}$-bound state. The main contribution is the transition from the $p$-scattering state to the $2s_{1/2}$-bound state (Table \ref{tab1}).

\begin{figure}[b]
    \includegraphics[width=1.\linewidth]{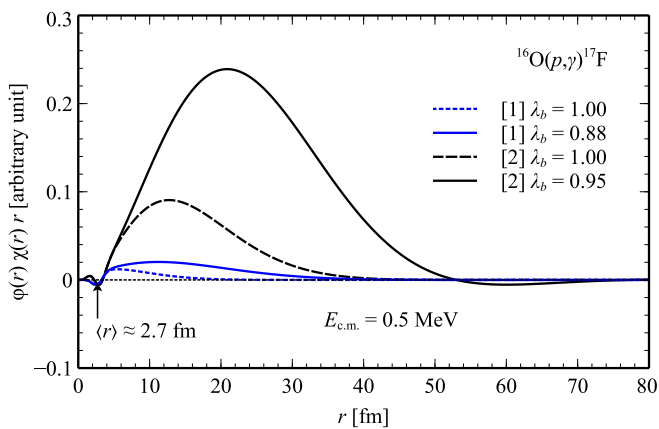}
    \caption{The radial dependence of overlap integrals for the $p$-scattering wave to the $2s_{1/2}$-bound states at $E_p = 0.5$ MeV (in c.m. frame). The transition to the ground state is noted as [1], and to the excited state [2].  The radius of $^{16}$O obtained from our HF calculation is $2.7$ fm. The calculations with and without the asymptotic correction are shown.}
    \label{overlapfig1}
\end{figure}

The correct asymptotic forms of the single-particle states $1d_{5/2}$ and $2s_{1/2}$ are important in the study of keV-nucleon RC reaction. Because of not only the extremely low energy but also the Coulomb and centrifugal barriers, the capture process occurs at a large distance. The overlap integral ($E_p = 0.5$ MeV) with and without the asymptotic correction are shown in Fig.~\ref{overlapfig1} for the transition to the ground state ([1]) and to the excited state ([2]). The overlap integral of the transition to the excited state is much larger than that of the transition to the ground state. It is a consequence of the halo properties of the bound $1/2^+$ state in $^{17}$F that was discussed in Refs.~\cite{mor97,ber03}. The proton-halo structure of the $1/2^+$ bound-state wave function is the cause of the increase of the $\mathcal{S}$ factor for the transition to the $1/2^+$ \cite{mor97}. Besides the astrophysics aspect, $^{17}$F is therefore a good candidate for the study of the weakly bound state as well as the proton-halo structure due to small separation energy, $Q = 0.11$ MeV.

The astrophysical $\mathcal{S}$ factors of two cases of $^{16}$O($p,\gamma$) have different trends when the energy decreases to zero. Due to the stronger centrifugal barrier for the $d$-wave, the astrophysical $\mathcal{S}$ factor of the transition to the ground state decreases. On the contrary, the $\mathcal{S}$ factor of the transition to the excited state raises. It is important for the extrapolation used in astrophysical studies.

\begin{figure}[b]
    \includegraphics[width=1.\linewidth]{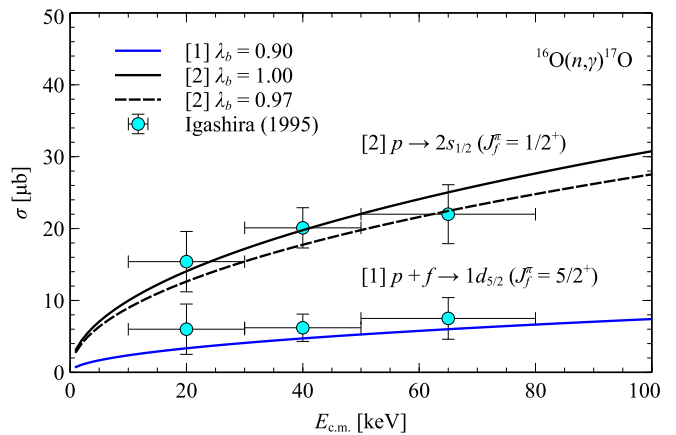}
    \caption{Calculated cross-section of the $^{16}$O($n,\gamma$)$^{17}$O reaction. [1] is for the transition to the ground state, and [2] is for the excited state. In [2], the calculations with $\lambda_b = 1.00$ and with the adjusted $\lambda_b = 0.97$ were shown for comparison ($S_F = 1.00$ for both cases). Experimental data were taken from Ref.~\cite{iga95}.}
    \label{16Ong}
\end{figure}

The study of keV-neutron RC by $^{16}$O was analyzed in Ref.~\cite{kit02} using the microscopic folding model. Our results are shown in Fig.~\ref{16Ong}. The transitions to the ground state ($5/2^+$) and the excited state ($1/2^+$) in $^{17}$O were both analyzed (Fig.~\ref{16Ong} and Table \ref{tab1}). Note that the factors $\lambda_b$ are close to unity as $\Delta e$ is close to zero. In addition, in the case of $^{16}$O($n,\gamma$)$^{17}$O$^{*}$, the transition to the excited state has exceptional property. The Coulomb and centrifugal barriers are both absent. It is possible to use the HF single-particle state obtained without the asymptotic correction. Indeed, the experimental data are also reasonably reproduced without the correction (the solid line in Fig.~\ref{16Ong}).

\subsection{$^{12}$C($p,\gamma$)$^{13}$N}
\begin{figure}[t]
    \includegraphics[width=1.\linewidth]{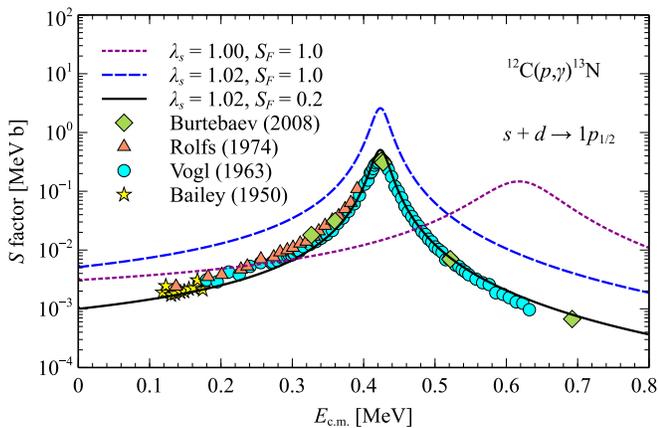}
    \caption{Calculated astrophysical $\mathcal{S}(E)$ of the $^{12}$C($p,\gamma$)$^{13}$N reaction. The calculations with $\lambda_s = 1.00$ and $\lambda_s = 1.02$ while keeping $S_F = 1.00$ are shown as the dotted and dashed line. The result with $S_F = 0.2$ is the solid line. Experimental data were taken from Refs.~\cite{bai50,bur08,rol74,vog63}.}
    \label{12Cpg}
\end{figure}

The HF calculation for $^{12}$C is not as good as in the case of double-magic nucleus $^{16}$O. However, with appropriate adjustments, the method of calculation is well applied for the study of RC from $^{12}$C and other light nuclei \cite{anh21PRC}. The $^{12}$C($p,\gamma$)$^{13}$N reaction is recalled because of the low-lying resonance at 0.42 MeV that was used as the calibration for the scattering wave function of both ($p,\gamma$) and ($n,\gamma$) in the study. To obtain the resonance at 0.42 MeV, the real part of central HF was slightly scaled by the factor $\lambda_s = 1.02$ (Fig.~\ref{12Cpg}). Note that, the continuum HF calculation gave a good description for the neutron scattering from $^{12}$C at low energy \cite{dov71} which implies that the spherical HF calculation is still available for $^{12}$C nucleus that is well-known deformed \cite{nak71} in some aspects.

The result of $^{12}$C($p,\gamma$)$^{13}$N reaction is shown in Fig.~\ref{12Cpg}. The direct capture to the ground state proceeds mainly via the $E1$ transitions from $s+d$-scattering states to the $1p_{1/2}$ bound state. The adjustments of $\lambda_s$ and $S_F$ are required in order to reproduce the resonance (Fig.~\ref{12Cpg}). It is expected that $S_F$ is now different from unity. In our calculation, it is $0.2$ to reproduce the experimental data. The nuclear shell-model calculation usually gives $S_F = 0.6$ for $1p_{1/2}$ single-particle state in $^{12}$C (Refs.~\cite{coh67,tim13} for example).

\subsection{$^{12}$C($n,\gamma$)$^{13}$C}
\begin{figure}[t]
    \includegraphics[width=1.\linewidth]{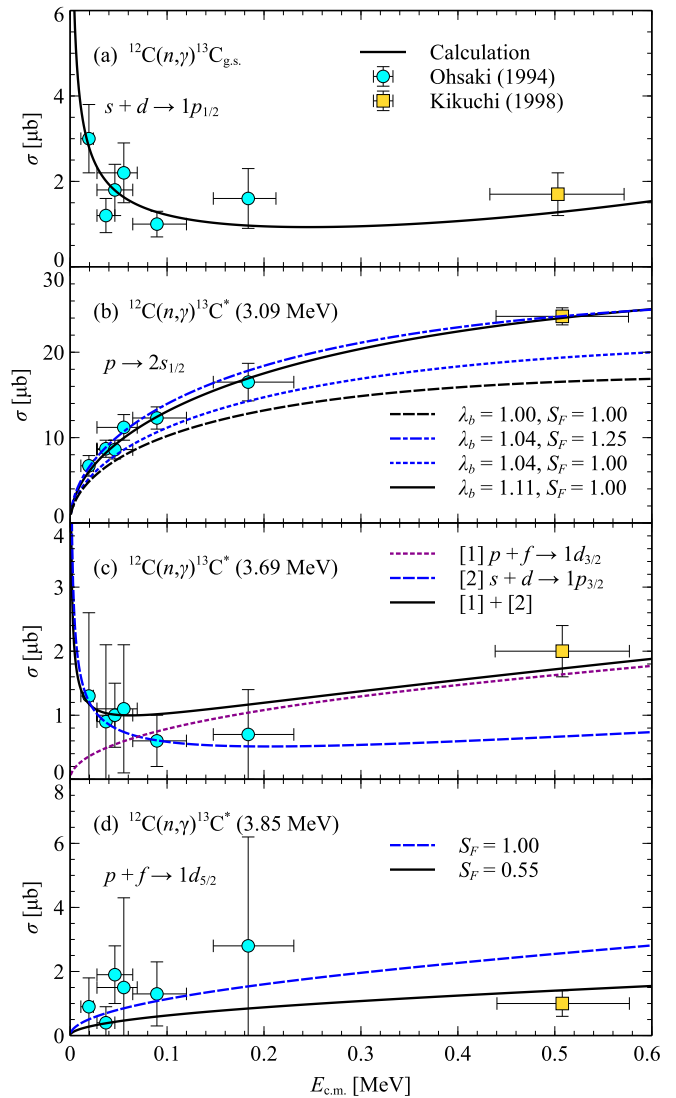}
    \caption{Calculated cross-sections of the ${}^{12}\mathrm{C}(n, \gamma){}^{13}\mathrm{C}$ reaction. The transitions to the ground state (a) and three excited states (b), (c), and (d) were analyzed. Experimental data were taken from Refs.~\cite{ohs94} (the circle), and \cite{kik98} (the square).}
    \label{12Cng}
\end{figure}

The scaling factor $\lambda_s$ is 1.02 for all nucleon RC reactions by $^{12}$C. The results of the capture cross-section for the first $E1$ transitions to the four bound states in $^{13}$C (ground state, and three excited states at 3.09 MeV, 3.69 MeV, and 3.85 MeV) are shown in Fig.~\ref{12Cng}. The parameters are given in Table \ref{tab1}. In the study of $^{12}$C($n,\gamma$) reaction, the experimental data points at 0.51 MeV in c.m. frame \cite{kik98} play the important role as similarly as low-lying resonances (Fig.~\ref{12Cng}). In the past, the study in Ref.~\cite{kit98} attempted to reproduce these data points with a microscopic approach.

In particular, in Fig.~\ref{12Cng}a, the calculated cross-section for the transition to the ground state in $^{13}$C ($1/2^-$) is in comparison with the measured cross-sections \cite{ohs94}. For low-energy neutrons, the $s$-scattering wave plays an important role in the transition to the ground state when the incident energy decreases. The factor $\lambda_b = 0.83$ that is close to $\lambda_b = 0.81$ of the $^{12}$C($p,\gamma$) reaction as the nuclear potentials in both reactions are almost the same.

The cross-sections for transition to the first excited state at 3.09 MeV in $^{13}$C$^*$ ($1/2^+$) are shown in Fig.~\ref{12Cng}b. The transition in this case is from the $p$-scattering state to the $2s_{1/2}$-bound state \cite{ots94,men95,kit98}. It seems to be successful to reproduce the experimental data (the dash-dotted line in Fig.~\ref{12Cng}b). The parameters are given in Table \ref{tab1}. However, the spectroscopic factor $S_F$ of $2s_{1/2}$ is $1.25$ in order to reach the data point at 0.51 MeV of Ref.~\cite{kik98}. With $S_F = 1.00$,
it is the dotted line in Fig.~\ref{12Cng}b.
In Ref.~\cite{kit98}, the value of $S_F$ was also approximately 1.25. Note that, the cross-section, in this case, is the largest, and the vertical error bar of the data point at 0.51 MeV in Fig.~\ref{12Cng}b is the smallest compared to the other cases in Fig.~\ref{12Cng}. To give a more physical value, the phenomenological Perey damping factor was introduced in the study of Ref.~\cite{kit98}, and the $S_F$ was then reduced to 0.95. However, the damping factor is already included microscopically within the Skyrme HF approximation in our calculation.

Our explanation for this case is as follows: firstly, the HF $2s_{1/2}$ single-particle state can be used as both Coulomb and centrifugal potentials are absent. The dashed line in Fig.~\ref{12Cng}b shows the calculation with $\lambda_b = 1.00$ that is significantly lower than the data point at 0.51 MeV. The reason is that the single-particle energy of $2s_{1/2}$ is $-1.25$ MeV according to our HF calculation. It is higher than the experimental neutron separation energy that is $-1.86$ MeV. It means that the neutron cannot be captured into this state as the system is not bound. To solve this problem, the $2s_{1/2}$ single-particle state was made to be more bound by adjusting the parameter $\lambda_b$. With $\lambda_b = 1.11$, the experimental data are well reproduced (the solid line in Fig.~\ref{12Cng}b). The single-particle energy of $2s_{1/2}$ is now $-3.02$ MeV, and the spectroscopic factor of $2s_{1/2}$ single-particle state is $S_F = 1.00$ (Table~\ref{tab1}).

Fig.~\ref{12Cng}c illustrates the comparison between observed cross-sections for the transitions to the $3/2^-$ state in $^{13}$C$^*$ at 3.69 MeV and theoretical cross-sections. In our HF calculation with the parameters were given in Table \ref{tab1}, the transition from $s+d$-scattering states to the $1p_{3/2}$-bound state does not seem possible as $1p_{3/2}$ is fully occupied. The transition from $p+f$-scattering states to the $1d_{3/2}$-bound state that is unoccupied is possible. It is shown in Fig.~\ref{12Cng}c that the experimental data are reproduced (the dotted line).

However, as the $p$-shell is not fully occupied in $^{12}$C, there is a possibility for the transition to $1p_{3/2}$-bound state. Moreover, it was proposed in Refs.~\cite{men95,kit98} that the transitions from $s+d$-scattering states to the $1p_{3/2}$-bound state is more physically correct. In our approach, even this complicated case was simplified by adjusting the parameter $\lambda_b$. The transition from $s+d$-scattering states to the $1p_{3/2}$-bound state in our calculation is shown in Fig.~\ref{12Cng}c (the dashed line). As the single-particle HF energy $\epsilon_{1p_{3/2}} = -16.85$ MeV that means $\Delta e = 15.59$ MeV, $\lambda_b$ has to be dramatically reduced to 0.45 in order to correct the asymptotic form. Although the data of Ref.~\cite{ohs94} is well-reproduced, it underestimates the data of Ref.~\cite{kik98}.

Our solution is the admixture of configuration that means both transitions contribute to the case. The result is the combination of transition to the $1d_{3/2}$ and to the $1p_{3/2}$ shown as the solid line in Fig.~\ref{12Cng}c. The spectroscopic factors from our analysis are reasonable. It is 0.25 for the $1p_{3/2}$ single-particle state. The $S_F$ value of 0.19 was given in the shell-model calculation \cite{coh67}. Experimentally, the values of 0.26 and 0.14 were reported in Refs.~\cite{tak77} and \cite{ohn86}, respectively. The spectroscopic factor of the $1d_{3/2}$ single-particle state is 0.95.

In Fig.~\ref{12Cng}d, the cross-sections for the transition leading to the $1d_{5/2}$ in the $5/2^+$ bound state of $^{13}$C$^{*}$ at 3.85 MeV are shown. The parameters were given in Table~\ref{tab1}. In this case, there are the $p$- and $f$-scattering waves. The dominant contribution is caused by the $p$-scattering wave. The result with $S_F = 1.00$ overestimates the data of Ref.~\cite{kik98}. Therefore, it is reduced to 0.55. The experimental value of $S_F$ is 0.58 as reported in Ref.~\cite{ohn86}.

When the reaction rate is of interest, the sum of all cross-sections from Figs.~\ref{12Cng}a-d is important, not the individual. At the energy lower than $300$ keV, the main contribution is caused by the transition with the $s$-scattering wave captured into the $p$-bound state (Fig.~\ref{12Cng}a). At higher energies ($E > 300$ keV), however, the capture is dominated by the transition from the $p$-scattering wave captured into the $s$- or $d$-bound state (Figs.~\ref{12Cng}b and \ref{12Cng}d). Note that it can be observed in Fig.~\ref{12Cng}c as both transitions are present.

\section{CONCLUSIONS} \label{sec4}
The bound-to-continuum potential model was successfully applied to reproduce the cross-sections of keV-nucleon RC reactions. The HF calculation provides the essentials for all calculations in the study. The asymptotic form of the single-particle bound state in the study of RC reaction is simply taken into account by the well-depth method. The value of spectroscopic factors $S_F$ and the properties of single-particle states are in agreement with experimental and theoretical studies. Although the theoretical calculations gave good predictions, experimental studies such as Ref.~\cite{kik98} are vital in the subject. Finally, the nucleon RC reactions are useful for the study of the interplay of quantum many-body physics and the dynamics of nuclear reactions as scattering states and weakly bound states are involved and can be treated simultaneously.

\section*{Acknowledgement}
We would like to thank Prof. N. Auerbach and Prof. V. Zelevinsky for their discussions. The present work is supported by Vietnam National Foundation for Science and Technology Development (NAFOSTED).

\bibliography{refs}

\begin{thebibliography}{52}%
\makeatletter
\providecommand \@ifxundefined [1]{%
 \@ifx{#1\undefined}
}%
\providecommand \@ifnum [1]{%
 \ifnum #1\expandafter \@firstoftwo
 \else \expandafter \@secondoftwo
 \fi
}%
\providecommand \@ifx [1]{%
 \ifx #1\expandafter \@firstoftwo
 \else \expandafter \@secondoftwo
 \fi
}%
\providecommand \natexlab [1]{#1}%
\providecommand \enquote  [1]{``#1''}%
\providecommand \bibnamefont  [1]{#1}%
\providecommand \bibfnamefont [1]{#1}%
\providecommand \citenamefont [1]{#1}%
\providecommand \href@noop [0]{\@secondoftwo}%
\providecommand \href [0]{\begingroup \@sanitize@url \@href}%
\providecommand \@href[1]{\@@startlink{#1}\@@href}%
\providecommand \@@href[1]{\endgroup#1\@@endlink}%
\providecommand \@sanitize@url [0]{\catcode `\\12\catcode `\$12\catcode
  `\&12\catcode `\#12\catcode `\^12\catcode `\_12\catcode `\%12\relax}%
\providecommand \@@startlink[1]{}%
\providecommand \@@endlink[0]{}%
\providecommand \url  [0]{\begingroup\@sanitize@url \@url }%
\providecommand \@url [1]{\endgroup\@href {#1}{\urlprefix }}%
\providecommand \urlprefix  [0]{URL }%
\providecommand \Eprint [0]{\href }%
\providecommand \doibase [0]{https://doi.org/}%
\providecommand \selectlanguage [0]{\@gobble}%
\providecommand \bibinfo  [0]{\@secondoftwo}%
\providecommand \bibfield  [0]{\@secondoftwo}%
\providecommand \translation [1]{[#1]}%
\providecommand \BibitemOpen [0]{}%
\providecommand \bibitemStop [0]{}%
\providecommand \bibitemNoStop [0]{.\EOS\space}%
\providecommand \EOS [0]{\spacefactor3000\relax}%
\providecommand \BibitemShut  [1]{\csname bibitem#1\endcsname}%
\let\auto@bib@innerbib\@empty
\bibitem [{\citenamefont {Rolfs}\ and\ \citenamefont {Barnes}(1990)}]{rol90}%
  \BibitemOpen
  \bibfield  {author} {\bibinfo {author} {\bibfnamefont {C.}~\bibnamefont
  {Rolfs}}\ and\ \bibinfo {author} {\bibfnamefont {C.~A.}\ \bibnamefont
  {Barnes}},\ }\bibfield  {title} {\bibinfo {title} {{Radiative Capture
  Reactions in Nuclear Astrophysics}},\ }\href
  {https://doi.org/10.1146/annurev.ns.40.120190.000401} {\bibfield  {journal}
  {\bibinfo  {journal} {Annu. Rev. Nucl. Part. Sci.}\ }\textbf {\bibinfo
  {volume} {40}},\ \bibinfo {pages} {45} (\bibinfo {year} {1990})}\BibitemShut
  {NoStop}%
\bibitem [{\citenamefont {Brune}\ and\ \citenamefont {Davids}(2015)}]{bru15}%
  \BibitemOpen
  \bibfield  {author} {\bibinfo {author} {\bibfnamefont {C.~R.}\ \bibnamefont
  {Brune}}\ and\ \bibinfo {author} {\bibfnamefont {B.}~\bibnamefont {Davids}},\
  }\bibfield  {title} {\bibinfo {title} {Radiative capture reactions in
  astrophysics},\ }\href {https://doi.org/10.1146/annurev-nucl-102014-022027}
  {\bibfield  {journal} {\bibinfo  {journal} {Annu. Rev. Nucl. Part. Sci.}\
  }\textbf {\bibinfo {volume} {65}},\ \bibinfo {pages} {87} (\bibinfo {year}
  {2015})}\BibitemShut {NoStop}%
\bibitem [{\citenamefont {Nunes}\ \emph {et~al.}(2020)\citenamefont {Nunes},
  \citenamefont {Potel}, \citenamefont {Poxon-Pearson},\ and\ \citenamefont
  {Cizewski}}]{nun20}%
  \BibitemOpen
  \bibfield  {author} {\bibinfo {author} {\bibfnamefont {F.~M.}\ \bibnamefont
  {Nunes}}, \bibinfo {author} {\bibfnamefont {G.}~\bibnamefont {Potel}},
  \bibinfo {author} {\bibfnamefont {T.}~\bibnamefont {Poxon-Pearson}},\ and\
  \bibinfo {author} {\bibfnamefont {J.~A.}\ \bibnamefont {Cizewski}},\
  }\bibfield  {title} {\bibinfo {title} {{Nuclear Reactions in Astrophysics: A
  Review of Useful Probes for Extracting Reaction Rates}},\ }\href
  {https://doi.org/10.1146/annurev-nucl-020620-063734} {\bibfield  {journal}
  {\bibinfo  {journal} {Annu. Rev. Nucl. Part. Sci.}\ }\textbf {\bibinfo
  {volume} {70}},\ \bibinfo {pages} {147} (\bibinfo {year} {2020})}\BibitemShut
  {NoStop}%
\bibitem [{\citenamefont {Descouvemont}(2020)}]{des20}%
  \BibitemOpen
  \bibfield  {author} {\bibinfo {author} {\bibfnamefont {P.}~\bibnamefont
  {Descouvemont}},\ }\bibfield  {title} {\bibinfo {title} {{Nuclear Reactions
  of Astrophysical Interest}},\ }\href
  {https://doi.org/10.3389/fspas.2020.00009} {\bibfield  {journal} {\bibinfo
  {journal} {Front. Astron. Space Sci.}\ }\textbf {\bibinfo {volume} {7}},\
  \bibinfo {pages} {9} (\bibinfo {year} {2020})}\BibitemShut {NoStop}%
\bibitem [{\citenamefont {Burbidge}\ \emph {et~al.}(1957)\citenamefont
  {Burbidge}, \citenamefont {Burbidge}, \citenamefont {Fowler},\ and\
  \citenamefont {Hoyle}}]{bur57}%
  \BibitemOpen
  \bibfield  {author} {\bibinfo {author} {\bibfnamefont {E.~M.}\ \bibnamefont
  {Burbidge}}, \bibinfo {author} {\bibfnamefont {G.~R.}\ \bibnamefont
  {Burbidge}}, \bibinfo {author} {\bibfnamefont {W.~A.}\ \bibnamefont
  {Fowler}},\ and\ \bibinfo {author} {\bibfnamefont {F.}~\bibnamefont
  {Hoyle}},\ }\bibfield  {title} {\bibinfo {title} {{Synthesis of the Elements
  in Stars}},\ }\href {https://doi.org/10.1103/RevModPhys.29.547} {\bibfield
  {journal} {\bibinfo  {journal} {Rev. Mod. Phys.}\ }\textbf {\bibinfo {volume}
  {29}},\ \bibinfo {pages} {547} (\bibinfo {year} {1957})}\BibitemShut
  {NoStop}%
\bibitem [{\citenamefont {Arcones}\ \emph {et~al.}(2017)\citenamefont
  {Arcones}, \citenamefont {Bardayan}, \citenamefont {Beers}, \citenamefont
  {Bernstein}, \citenamefont {Blackmon}, \citenamefont {Messer}, \citenamefont
  {Brown}, \citenamefont {Brown}, \citenamefont {Brune}, \citenamefont
  {Champagne}, \citenamefont {Chieffi}, \citenamefont {Couture}, \citenamefont
  {Danielewicz}, \citenamefont {Diehl}, \citenamefont {El-Eid}, \citenamefont
  {Escher}, \citenamefont {Fields}, \citenamefont {Fröhlich}, \citenamefont
  {Herwig}, \citenamefont {Hix}, \citenamefont {Iliadis}, \citenamefont
  {Lynch}, \citenamefont {McLaughlin}, \citenamefont {Meyer}, \citenamefont
  {Mezzacappa}, \citenamefont {Nunes}, \citenamefont {O’Shea}, \citenamefont
  {Prakash}, \citenamefont {Pritychenko}, \citenamefont {Reddy}, \citenamefont
  {Rehm}, \citenamefont {Rogachev}, \citenamefont {Rutledge}, \citenamefont
  {Schatz}, \citenamefont {Smith}, \citenamefont {Stairs}, \citenamefont
  {Steiner}, \citenamefont {Strohmayer}, \citenamefont {Timmes}, \citenamefont
  {Townsley}, \citenamefont {Wiescher}, \citenamefont {Zegers},\ and\
  \citenamefont {Zingale}}]{arc17}%
  \BibitemOpen
  \bibfield  {author} {\bibinfo {author} {\bibfnamefont {A.}~\bibnamefont
  {Arcones}}, \bibinfo {author} {\bibfnamefont {D.~W.}\ \bibnamefont
  {Bardayan}}, \bibinfo {author} {\bibfnamefont {T.~C.}\ \bibnamefont {Beers}},
  \bibinfo {author} {\bibfnamefont {L.~A.}\ \bibnamefont {Bernstein}}, \bibinfo
  {author} {\bibfnamefont {J.~C.}\ \bibnamefont {Blackmon}}, \bibinfo {author}
  {\bibfnamefont {B.}~\bibnamefont {Messer}}, \bibinfo {author} {\bibfnamefont
  {B.~A.}\ \bibnamefont {Brown}}, \bibinfo {author} {\bibfnamefont {E.~F.}\
  \bibnamefont {Brown}}, \bibinfo {author} {\bibfnamefont {C.~R.}\ \bibnamefont
  {Brune}}, \bibinfo {author} {\bibfnamefont {A.~E.}\ \bibnamefont
  {Champagne}}, \bibinfo {author} {\bibfnamefont {A.}~\bibnamefont {Chieffi}},
  \bibinfo {author} {\bibfnamefont {A.~J.}\ \bibnamefont {Couture}}, \bibinfo
  {author} {\bibfnamefont {P.}~\bibnamefont {Danielewicz}}, \bibinfo {author}
  {\bibfnamefont {R.}~\bibnamefont {Diehl}}, \bibinfo {author} {\bibfnamefont
  {M.}~\bibnamefont {El-Eid}}, \bibinfo {author} {\bibfnamefont {J.~E.}\
  \bibnamefont {Escher}}, \bibinfo {author} {\bibfnamefont {B.~D.}\
  \bibnamefont {Fields}}, \bibinfo {author} {\bibfnamefont {C.}~\bibnamefont
  {Fröhlich}}, \bibinfo {author} {\bibfnamefont {F.}~\bibnamefont {Herwig}},
  \bibinfo {author} {\bibfnamefont {W.~R.}\ \bibnamefont {Hix}}, \bibinfo
  {author} {\bibfnamefont {C.}~\bibnamefont {Iliadis}}, \bibinfo {author}
  {\bibfnamefont {W.~G.}\ \bibnamefont {Lynch}}, \bibinfo {author}
  {\bibfnamefont {G.~C.}\ \bibnamefont {McLaughlin}}, \bibinfo {author}
  {\bibfnamefont {B.~S.}\ \bibnamefont {Meyer}}, \bibinfo {author}
  {\bibfnamefont {A.}~\bibnamefont {Mezzacappa}}, \bibinfo {author}
  {\bibfnamefont {F.}~\bibnamefont {Nunes}}, \bibinfo {author} {\bibfnamefont
  {B.~W.}\ \bibnamefont {O’Shea}}, \bibinfo {author} {\bibfnamefont
  {M.}~\bibnamefont {Prakash}}, \bibinfo {author} {\bibfnamefont
  {B.}~\bibnamefont {Pritychenko}}, \bibinfo {author} {\bibfnamefont
  {S.}~\bibnamefont {Reddy}}, \bibinfo {author} {\bibfnamefont
  {E.}~\bibnamefont {Rehm}}, \bibinfo {author} {\bibfnamefont {G.}~\bibnamefont
  {Rogachev}}, \bibinfo {author} {\bibfnamefont {R.~E.}\ \bibnamefont
  {Rutledge}}, \bibinfo {author} {\bibfnamefont {H.}~\bibnamefont {Schatz}},
  \bibinfo {author} {\bibfnamefont {M.~S.}\ \bibnamefont {Smith}}, \bibinfo
  {author} {\bibfnamefont {I.~H.}\ \bibnamefont {Stairs}}, \bibinfo {author}
  {\bibfnamefont {A.~W.}\ \bibnamefont {Steiner}}, \bibinfo {author}
  {\bibfnamefont {T.~E.}\ \bibnamefont {Strohmayer}}, \bibinfo {author}
  {\bibfnamefont {F.}~\bibnamefont {Timmes}}, \bibinfo {author} {\bibfnamefont
  {D.~M.}\ \bibnamefont {Townsley}}, \bibinfo {author} {\bibfnamefont
  {M.}~\bibnamefont {Wiescher}}, \bibinfo {author} {\bibfnamefont {R.~G.}\
  \bibnamefont {Zegers}},\ and\ \bibinfo {author} {\bibfnamefont
  {M.}~\bibnamefont {Zingale}},\ }\bibfield  {title} {\bibinfo {title} {White
  paper on nuclear astrophysics and low energy nuclear physics part 1: Nuclear
  astrophysics},\ }\href {https://doi.org/doi.org/10.1016/j.ppnp.2016.12.003}
  {\bibfield  {journal} {\bibinfo  {journal} {Prog. Part. Nucl. Phys.}\
  }\textbf {\bibinfo {volume} {94}},\ \bibinfo {pages} {1} (\bibinfo {year}
  {2017})}\BibitemShut {NoStop}%
\bibitem [{\citenamefont {Villante}\ and\ \citenamefont
  {Serenelli}(2021)}]{vil21}%
  \BibitemOpen
  \bibfield  {author} {\bibinfo {author} {\bibfnamefont {F.~L.}\ \bibnamefont
  {Villante}}\ and\ \bibinfo {author} {\bibfnamefont {A.}~\bibnamefont
  {Serenelli}},\ }\bibfield  {title} {\bibinfo {title} {{The relevance of
  nuclear reactions for Standard Solar Models construction}},\ }\href
  {https://doi.org/10.3389/fspas.2020.618356} {\bibfield  {journal} {\bibinfo
  {journal} {Front. Astron. Space Sci.}\ }\textbf {\bibinfo {volume} {7}},\
  \bibinfo {pages} {112} (\bibinfo {year} {2021})}\BibitemShut {NoStop}%
\bibitem [{\citenamefont {Oberhummer}\ \emph {et~al.}(1996)\citenamefont
  {Oberhummer}, \citenamefont {Herndl}, \citenamefont {Rauscher},\ and\
  \citenamefont {Beer}}]{obe96}%
  \BibitemOpen
  \bibfield  {author} {\bibinfo {author} {\bibfnamefont {H.}~\bibnamefont
  {Oberhummer}}, \bibinfo {author} {\bibfnamefont {H.}~\bibnamefont {Herndl}},
  \bibinfo {author} {\bibfnamefont {T.}~\bibnamefont {Rauscher}},\ and\
  \bibinfo {author} {\bibfnamefont {H.}~\bibnamefont {Beer}},\ }\bibfield
  {title} {\bibinfo {title} {Neutron-induced nucleosynthesis},\ }\href
  {https://doi.org/10.1007/BF01931785} {\bibfield  {journal} {\bibinfo
  {journal} {Surv. Geophys.}\ }\textbf {\bibinfo {volume} {17}},\ \bibinfo
  {pages} {665} (\bibinfo {year} {1996})}\BibitemShut {NoStop}%
\bibitem [{\citenamefont {Kajino}(2002)}]{kaj02}%
  \BibitemOpen
  \bibfield  {author} {\bibinfo {author} {\bibfnamefont {T.}~\bibnamefont
  {Kajino}},\ }\bibfield  {title} {\bibinfo {title} {{Cosmological
  Nucleosynthesis in the Big-Bang and Supernovae}},\ }\href
  {https://doi.org/10.1080/00223131.2002.10875156} {\bibfield  {journal}
  {\bibinfo  {journal} {J. Nucl. Sci. Technol.}\ }\textbf {\bibinfo {volume}
  {39}},\ \bibinfo {pages} {530} (\bibinfo {year} {2002})}\BibitemShut
  {NoStop}%
\bibitem [{\citenamefont {Busso}\ \emph {et~al.}(1999)\citenamefont {Busso},
  \citenamefont {Gallino},\ and\ \citenamefont {Wasserburg}}]{bus99}%
  \BibitemOpen
  \bibfield  {author} {\bibinfo {author} {\bibfnamefont {M.}~\bibnamefont
  {Busso}}, \bibinfo {author} {\bibfnamefont {R.}~\bibnamefont {Gallino}},\
  and\ \bibinfo {author} {\bibfnamefont {G.~J.}\ \bibnamefont {Wasserburg}},\
  }\bibfield  {title} {\bibinfo {title} {{Nucleosynthesis in {A}symptotic
  {G}iant {B}ranch Stars: {R}elevance for Galactic Enrichment and Solar System
  Formation}},\ }\href {https://doi.org/10.1146/annurev.astro.37.1.239}
  {\bibfield  {journal} {\bibinfo  {journal} {Annu. Rev. Astron. Astrophys.}\
  }\textbf {\bibinfo {volume} {37}},\ \bibinfo {pages} {239} (\bibinfo {year}
  {1999})}\BibitemShut {NoStop}%
\bibitem [{\citenamefont {Mohr}\ \emph {et~al.}(2016)\citenamefont {Mohr},
  \citenamefont {Heinz}, \citenamefont {Pignatari}, \citenamefont {Dillmann},
  \citenamefont {Mengoni},\ and\ \citenamefont {K{\"a}ppeler}}]{moh16}%
  \BibitemOpen
  \bibfield  {author} {\bibinfo {author} {\bibfnamefont {P.}~\bibnamefont
  {Mohr}}, \bibinfo {author} {\bibfnamefont {C.}~\bibnamefont {Heinz}},
  \bibinfo {author} {\bibfnamefont {M.}~\bibnamefont {Pignatari}}, \bibinfo
  {author} {\bibfnamefont {I.}~\bibnamefont {Dillmann}}, \bibinfo {author}
  {\bibfnamefont {A.}~\bibnamefont {Mengoni}},\ and\ \bibinfo {author}
  {\bibfnamefont {F.}~\bibnamefont {K{\"a}ppeler}},\ }\bibfield  {title}
  {\bibinfo {title} {Re-evaluation of the $^{16}\mathrm{O}(n,
  \gamma)^{17}\mathrm{O}$ cross section at astrophysical energies and its role
  as a neutron poison in the $s$-process},\ }\href
  {https://doi.org/10.3847/0004-637X/827/1/29} {\bibfield  {journal} {\bibinfo
  {journal} {Astrophys. J.}\ }\textbf {\bibinfo {volume} {827}},\ \bibinfo
  {pages} {29} (\bibinfo {year} {2016})}\BibitemShut {NoStop}%
\bibitem [{\citenamefont {Carlson}\ \emph {et~al.}(2017)\citenamefont
  {Carlson}, \citenamefont {Carpenter}, \citenamefont {Casten}, \citenamefont
  {Elster}, \citenamefont {Fallon}, \citenamefont {Gade}, \citenamefont
  {Gross}, \citenamefont {Hagen}, \citenamefont {Hayes}, \citenamefont
  {Higinbotham}, \citenamefont {Howell}, \citenamefont {Horowitz},
  \citenamefont {Jones}, \citenamefont {Kondev}, \citenamefont {Lapi},
  \citenamefont {Macchiavelli}, \citenamefont {McCutchen}, \citenamefont
  {Natowitz}, \citenamefont {Nazarewicz}, \citenamefont {Papenbrock},
  \citenamefont {Reddy}, \citenamefont {Riley}, \citenamefont {Savage},
  \citenamefont {Savard}, \citenamefont {Sherrill}, \citenamefont {Sobotka},
  \citenamefont {Stoyer}, \citenamefont {{Betty Tsang}}, \citenamefont
  {Vetter}, \citenamefont {Wiedenhoever}, \citenamefont {Wuosmaa},\ and\
  \citenamefont {Yennello}}]{car17}%
  \BibitemOpen
  \bibfield  {author} {\bibinfo {author} {\bibfnamefont {J.}~\bibnamefont
  {Carlson}}, \bibinfo {author} {\bibfnamefont {M.~P.}\ \bibnamefont
  {Carpenter}}, \bibinfo {author} {\bibfnamefont {R.}~\bibnamefont {Casten}},
  \bibinfo {author} {\bibfnamefont {C.}~\bibnamefont {Elster}}, \bibinfo
  {author} {\bibfnamefont {P.}~\bibnamefont {Fallon}}, \bibinfo {author}
  {\bibfnamefont {A.}~\bibnamefont {Gade}}, \bibinfo {author} {\bibfnamefont
  {C.}~\bibnamefont {Gross}}, \bibinfo {author} {\bibfnamefont
  {G.}~\bibnamefont {Hagen}}, \bibinfo {author} {\bibfnamefont {A.~C.}\
  \bibnamefont {Hayes}}, \bibinfo {author} {\bibfnamefont {D.~W.}\ \bibnamefont
  {Higinbotham}}, \bibinfo {author} {\bibfnamefont {C.~R.}\ \bibnamefont
  {Howell}}, \bibinfo {author} {\bibfnamefont {C.~J.}\ \bibnamefont
  {Horowitz}}, \bibinfo {author} {\bibfnamefont {K.~L.}\ \bibnamefont {Jones}},
  \bibinfo {author} {\bibfnamefont {F.~G.}\ \bibnamefont {Kondev}}, \bibinfo
  {author} {\bibfnamefont {S.}~\bibnamefont {Lapi}}, \bibinfo {author}
  {\bibfnamefont {A.}~\bibnamefont {Macchiavelli}}, \bibinfo {author}
  {\bibfnamefont {E.~A.}\ \bibnamefont {McCutchen}}, \bibinfo {author}
  {\bibfnamefont {J.}~\bibnamefont {Natowitz}}, \bibinfo {author}
  {\bibfnamefont {W.}~\bibnamefont {Nazarewicz}}, \bibinfo {author}
  {\bibfnamefont {T.}~\bibnamefont {Papenbrock}}, \bibinfo {author}
  {\bibfnamefont {S.}~\bibnamefont {Reddy}}, \bibinfo {author} {\bibfnamefont
  {M.~A.}\ \bibnamefont {Riley}}, \bibinfo {author} {\bibfnamefont {M.~J.}\
  \bibnamefont {Savage}}, \bibinfo {author} {\bibfnamefont {G.}~\bibnamefont
  {Savard}}, \bibinfo {author} {\bibfnamefont {B.~M.}\ \bibnamefont
  {Sherrill}}, \bibinfo {author} {\bibfnamefont {L.~G.}\ \bibnamefont
  {Sobotka}}, \bibinfo {author} {\bibfnamefont {M.~A.}\ \bibnamefont {Stoyer}},
  \bibinfo {author} {\bibfnamefont {M.}~\bibnamefont {{Betty Tsang}}}, \bibinfo
  {author} {\bibfnamefont {K.}~\bibnamefont {Vetter}}, \bibinfo {author}
  {\bibfnamefont {I.}~\bibnamefont {Wiedenhoever}}, \bibinfo {author}
  {\bibfnamefont {A.~H.}\ \bibnamefont {Wuosmaa}},\ and\ \bibinfo {author}
  {\bibfnamefont {S.}~\bibnamefont {Yennello}},\ }\bibfield  {title} {\bibinfo
  {title} {White paper on nuclear astrophysics and low-energy nuclear physics,
  part 2: Low-energy nuclear physics},\ }\href
  {https://doi.org/https://doi.org/10.1016/j.ppnp.2016.11.002} {\bibfield
  {journal} {\bibinfo  {journal} {Prog. Part. Nucl. Phys.}\ }\textbf {\bibinfo
  {volume} {94}},\ \bibinfo {pages} {68} (\bibinfo {year} {2017})}\BibitemShut
  {NoStop}%
\bibitem [{\citenamefont {Ho}\ and\ \citenamefont {Lone}(1983)}]{ho83}%
  \BibitemOpen
  \bibfield  {author} {\bibinfo {author} {\bibfnamefont {Y.-K.}\ \bibnamefont
  {Ho}}\ and\ \bibinfo {author} {\bibfnamefont {M.~A.}\ \bibnamefont {Lone}},\
  }\bibfield  {title} {\bibinfo {title} {{An interference effect in the channel
  radiative neutron capture process}},\ }\href
  {https://doi.org/https://doi.org/10.1016/0375-9474(83)90282-8} {\bibfield
  {journal} {\bibinfo  {journal} {Nucl. Phys. A}\ }\textbf {\bibinfo {volume}
  {406}},\ \bibinfo {pages} {18} (\bibinfo {year} {1983})}\BibitemShut
  {NoStop}%
\bibitem [{\citenamefont {Ho}\ \emph {et~al.}(1991)\citenamefont {Ho},
  \citenamefont {Kitazawa},\ and\ \citenamefont {Igashira}}]{ho91}%
  \BibitemOpen
  \bibfield  {author} {\bibinfo {author} {\bibfnamefont {Y.-K.}\ \bibnamefont
  {Ho}}, \bibinfo {author} {\bibfnamefont {H.}~\bibnamefont {Kitazawa}},\ and\
  \bibinfo {author} {\bibfnamefont {M.}~\bibnamefont {Igashira}},\ }\bibfield
  {title} {\bibinfo {title} {{Channel-capture mechanism in low-energy neutron
  capture by $^{12}\mathrm{C}$}},\ }\href
  {https://doi.org/10.1103/PhysRevC.44.1148} {\bibfield  {journal} {\bibinfo
  {journal} {Phys. Rev. C}\ }\textbf {\bibinfo {volume} {44}},\ \bibinfo
  {pages} {1148} (\bibinfo {year} {1991})}\BibitemShut {NoStop}%
\bibitem [{\citenamefont {Baye}(2004)}]{bay04}%
  \BibitemOpen
  \bibfield  {author} {\bibinfo {author} {\bibfnamefont {D.}~\bibnamefont
  {Baye}},\ }\bibfield  {title} {\bibinfo {title} {Cross section expansion for
  direct neutron radiative capture},\ }\href
  {https://doi.org/10.1103/PhysRevC.70.015801} {\bibfield  {journal} {\bibinfo
  {journal} {Phys. Rev. C}\ }\textbf {\bibinfo {volume} {70}},\ \bibinfo
  {pages} {015801} (\bibinfo {year} {2004})}\BibitemShut {NoStop}%
\bibitem [{\citenamefont {Huang}\ \emph {et~al.}(2010)\citenamefont {Huang},
  \citenamefont {Bertulani},\ and\ \citenamefont {Guimarães}}]{hua10}%
  \BibitemOpen
  \bibfield  {author} {\bibinfo {author} {\bibfnamefont {J.~T.}\ \bibnamefont
  {Huang}}, \bibinfo {author} {\bibfnamefont {C.~A.}\ \bibnamefont
  {Bertulani}},\ and\ \bibinfo {author} {\bibfnamefont {V.}~\bibnamefont
  {Guimarães}},\ }\bibfield  {title} {\bibinfo {title} {{Radiative capture of
  nucleons at astrophysical energies with single-particle states}},\ }\href
  {https://doi.org/https://doi.org/10.1016/j.adt.2010.06.004} {\bibfield
  {journal} {\bibinfo  {journal} {At. Data Nucl. Data Tables}\ }\textbf
  {\bibinfo {volume} {96}},\ \bibinfo {pages} {824} (\bibinfo {year}
  {2010})}\BibitemShut {NoStop}%
\bibitem [{\citenamefont {Dubovichenko}\ \emph {et~al.}(2013)\citenamefont
  {Dubovichenko}, \citenamefont {Dzhazairov-Kakhramanov},\ and\ \citenamefont
  {Burkova}}]{dub13}%
  \BibitemOpen
  \bibfield  {author} {\bibinfo {author} {\bibfnamefont {S.}~\bibnamefont
  {Dubovichenko}}, \bibinfo {author} {\bibfnamefont {A.}~\bibnamefont
  {Dzhazairov-Kakhramanov}},\ and\ \bibinfo {author} {\bibfnamefont
  {N.}~\bibnamefont {Burkova}},\ }\bibfield  {title} {\bibinfo {title} {{The
  radiative neutron capture on ${}^{2}\mathrm{H}$, ${}^{6}\mathrm{Li}$,
  ${}^{7}\mathrm{Li}$, ${}^{12}\mathrm{C}$ and $^{13}\mathrm{C}$ at
  astrophysical energies}},\ }\href {https://doi.org/10.1142/S0218301313500286}
  {\bibfield  {journal} {\bibinfo  {journal} {Int. J. Mod. Phys. E}\ }\textbf
  {\bibinfo {volume} {22}},\ \bibinfo {pages} {1350028} (\bibinfo {year}
  {2013})}\BibitemShut {NoStop}%
\bibitem [{\citenamefont {Xu}\ \emph {et~al.}(2013)\citenamefont {Xu},
  \citenamefont {Takahashi}, \citenamefont {Goriely}, \citenamefont {Arnould},
  \citenamefont {Ohta},\ and\ \citenamefont {Utsunomiya}}]{xu13}%
  \BibitemOpen
  \bibfield  {author} {\bibinfo {author} {\bibfnamefont {Y.}~\bibnamefont
  {Xu}}, \bibinfo {author} {\bibfnamefont {K.}~\bibnamefont {Takahashi}},
  \bibinfo {author} {\bibfnamefont {S.}~\bibnamefont {Goriely}}, \bibinfo
  {author} {\bibfnamefont {M.}~\bibnamefont {Arnould}}, \bibinfo {author}
  {\bibfnamefont {M.}~\bibnamefont {Ohta}},\ and\ \bibinfo {author}
  {\bibfnamefont {H.}~\bibnamefont {Utsunomiya}},\ }\bibfield  {title}
  {\bibinfo {title} {{{NACRE II}: an update of the {NACRE} compilation of
  charged-particle-induced thermonuclear reaction rates for nuclei with mass
  number ${A}<16$}},\ }\href
  {https://doi.org/https://doi.org/10.1016/j.nuclphysa.2013.09.007} {\bibfield
  {journal} {\bibinfo  {journal} {Nucl. Phys. A}\ }\textbf {\bibinfo {volume}
  {918}},\ \bibinfo {pages} {61} (\bibinfo {year} {2013})}\BibitemShut
  {NoStop}%
\bibitem [{\citenamefont {Kitazawa}\ \emph {et~al.}(2002)\citenamefont
  {Kitazawa}, \citenamefont {Igashira}, \citenamefont {Ohsaki},\ and\
  \citenamefont {Matsushima}}]{kit02}%
  \BibitemOpen
  \bibfield  {author} {\bibinfo {author} {\bibfnamefont {H.}~\bibnamefont
  {Kitazawa}}, \bibinfo {author} {\bibfnamefont {M.}~\bibnamefont {Igashira}},
  \bibinfo {author} {\bibfnamefont {T.}~\bibnamefont {Ohsaki}},\ and\ \bibinfo
  {author} {\bibfnamefont {T.}~\bibnamefont {Matsushima}},\ }\bibfield  {title}
  {\bibinfo {title} {{Folding Model Potential for the Neutron Direct Capture of
  ${}^{12}\mathrm{C}$ and ${}^{16}\mathrm{O}$}},\ }\href
  {https://doi.org/10.1080/00223131.2002.10875219} {\bibfield  {journal}
  {\bibinfo  {journal} {J. Nucl. Sci. Technol.}\ }\textbf {\bibinfo {volume}
  {39}},\ \bibinfo {pages} {799} (\bibinfo {year} {2002})}\BibitemShut
  {NoStop}%
\bibitem [{\citenamefont {Anh}\ \emph {et~al.}(2021)\citenamefont {Anh},
  \citenamefont {Phuc}, \citenamefont {Khoa}, \citenamefont {Chien},\ and\
  \citenamefont {Phuc}}]{anh21NPA}%
  \BibitemOpen
  \bibfield  {author} {\bibinfo {author} {\bibfnamefont {N.~L.}\ \bibnamefont
  {Anh}}, \bibinfo {author} {\bibfnamefont {N.~H.}\ \bibnamefont {Phuc}},
  \bibinfo {author} {\bibfnamefont {D.~T.}\ \bibnamefont {Khoa}}, \bibinfo
  {author} {\bibfnamefont {L.~H.}\ \bibnamefont {Chien}},\ and\ \bibinfo
  {author} {\bibfnamefont {N.~T.~T.}\ \bibnamefont {Phuc}},\ }\bibfield
  {title} {\bibinfo {title} {{Folding model approach to the elastic
  $p+{}^{12,13}\mathrm{C}$ scattering at low energies and radiative capture
  ${}^{12,13}\mathrm{C}(p,\gamma)$ reactions}},\ }\href
  {https://doi.org/10.1016/j.nuclphysa.2020.122078} {\bibfield  {journal}
  {\bibinfo  {journal} {Nucl. Phys. A}\ }\textbf {\bibinfo {volume} {1006}},\
  \bibinfo {pages} {122078} (\bibinfo {year} {2021})}\BibitemShut {NoStop}%
\bibitem [{\citenamefont {Dufour}\ and\ \citenamefont
  {Descouvemont}(1997)}]{duf97}%
  \BibitemOpen
  \bibfield  {author} {\bibinfo {author} {\bibfnamefont {M.}~\bibnamefont
  {Dufour}}\ and\ \bibinfo {author} {\bibfnamefont {P.}~\bibnamefont
  {Descouvemont}},\ }\bibfield  {title} {\bibinfo {title} {{Multicluster study
  of the ${}^{12}\mathrm{C}+n$ and ${}^{12}\mathrm{C}+p$ systems}},\ }\href
  {https://doi.org/10.1103/PhysRevC.56.1831} {\bibfield  {journal} {\bibinfo
  {journal} {Phys. Rev. C}\ }\textbf {\bibinfo {volume} {56}},\ \bibinfo
  {pages} {1831} (\bibinfo {year} {1997})}\BibitemShut {NoStop}%
\bibitem [{\citenamefont {Anh}\ and\ \citenamefont {Loc}(2021)}]{anh21PRC}%
  \BibitemOpen
  \bibfield  {author} {\bibinfo {author} {\bibfnamefont {N.~L.}\ \bibnamefont
  {Anh}}\ and\ \bibinfo {author} {\bibfnamefont {B.~M.}\ \bibnamefont {Loc}},\
  }\bibfield  {title} {\bibinfo {title} {{Bound-to-continuum potential model
  for the ($p,\ensuremath{\gamma}$) reactions of the CNO nucleosynthesis
  cycle}},\ }\href {https://doi.org/10.1103/PhysRevC.103.035812} {\bibfield
  {journal} {\bibinfo  {journal} {Phys. Rev. C}\ }\textbf {\bibinfo {volume}
  {103}},\ \bibinfo {pages} {035812} (\bibinfo {year} {2021})}\BibitemShut
  {NoStop}%
\bibitem [{\citenamefont {Chabanat}\ \emph {et~al.}(1998)\citenamefont
  {Chabanat}, \citenamefont {Bonche}, \citenamefont {Haensel}, \citenamefont
  {Meyer},\ and\ \citenamefont {Schaeffer}}]{cha98}%
  \BibitemOpen
  \bibfield  {author} {\bibinfo {author} {\bibfnamefont {E.}~\bibnamefont
  {Chabanat}}, \bibinfo {author} {\bibfnamefont {P.}~\bibnamefont {Bonche}},
  \bibinfo {author} {\bibfnamefont {P.}~\bibnamefont {Haensel}}, \bibinfo
  {author} {\bibfnamefont {J.}~\bibnamefont {Meyer}},\ and\ \bibinfo {author}
  {\bibfnamefont {R.}~\bibnamefont {Schaeffer}},\ }\bibfield  {title} {\bibinfo
  {title} {{A {S}kyrme parametrization from subnuclear to neutron star
  densities {P}art {II}. {N}uclei far from stabilities}},\ }\href
  {https://doi.org/https://doi.org/10.1016/S0375-9474(98)00180-8} {\bibfield
  {journal} {\bibinfo  {journal} {Nucl. Phys. A}\ }\textbf {\bibinfo {volume}
  {635}},\ \bibinfo {pages} {231} (\bibinfo {year} {1998})}\BibitemShut
  {NoStop}%
\bibitem [{\citenamefont {Chandel}\ \emph {et~al.}(2003)\citenamefont
  {Chandel}, \citenamefont {Dhiman},\ and\ \citenamefont {Shyam}}]{cha03}%
  \BibitemOpen
  \bibfield  {author} {\bibinfo {author} {\bibfnamefont {S.~S.}\ \bibnamefont
  {Chandel}}, \bibinfo {author} {\bibfnamefont {S.~K.}\ \bibnamefont
  {Dhiman}},\ and\ \bibinfo {author} {\bibfnamefont {R.}~\bibnamefont
  {Shyam}},\ }\bibfield  {title} {\bibinfo {title} {Structure of
  $^{8}\mathrm{B}$ and astrophysical ${S}_{17}$ factor in {S}kyrme
  {H}artree-{F}ock theory},\ }\href
  {https://doi.org/10.1103/PhysRevC.68.054320} {\bibfield  {journal} {\bibinfo
  {journal} {Phys. Rev. C}\ }\textbf {\bibinfo {volume} {68}},\ \bibinfo
  {pages} {054320} (\bibinfo {year} {2003})}\BibitemShut {NoStop}%
\bibitem [{\citenamefont {Rolfs}(1973)}]{rol73}%
  \BibitemOpen
  \bibfield  {author} {\bibinfo {author} {\bibfnamefont {C.}~\bibnamefont
  {Rolfs}},\ }\bibfield  {title} {\bibinfo {title} {{Spectroscopic factors from
  radiative capture reactions}},\ }\href
  {https://doi.org/https://doi.org/10.1016/0375-9474(73)90622-2} {\bibfield
  {journal} {\bibinfo  {journal} {Nucl. Phys. A}\ }\textbf {\bibinfo {volume}
  {217}},\ \bibinfo {pages} {29 } (\bibinfo {year} {1973})}\BibitemShut
  {NoStop}%
\bibitem [{\citenamefont {Mengoni}\ \emph {et~al.}(1995)\citenamefont
  {Mengoni}, \citenamefont {Otsuka},\ and\ \citenamefont {Ishihara}}]{men95}%
  \BibitemOpen
  \bibfield  {author} {\bibinfo {author} {\bibfnamefont {A.}~\bibnamefont
  {Mengoni}}, \bibinfo {author} {\bibfnamefont {T.}~\bibnamefont {Otsuka}},\
  and\ \bibinfo {author} {\bibfnamefont {M.}~\bibnamefont {Ishihara}},\
  }\bibfield  {title} {\bibinfo {title} {{Direct radiative capture of $p$-wave
  neutrons}},\ }\href {https://doi.org/10.1103/PhysRevC.52.R2334} {\bibfield
  {journal} {\bibinfo  {journal} {Phys. Rev. C}\ }\textbf {\bibinfo {volume}
  {52}},\ \bibinfo {pages} {R2334} (\bibinfo {year} {1995})}\BibitemShut
  {NoStop}%
\bibitem [{\citenamefont {Dover}\ and\ \citenamefont {Giai}(1971)}]{dov71}%
  \BibitemOpen
  \bibfield  {author} {\bibinfo {author} {\bibfnamefont {C.~B.}\ \bibnamefont
  {Dover}}\ and\ \bibinfo {author} {\bibfnamefont {N.~V.}\ \bibnamefont
  {Giai}},\ }\bibfield  {title} {\bibinfo {title} {{Low-energy neutron
  scattering by a {H}artree-{F}ock field}},\ }\href
  {https://doi.org/https://doi.org/10.1016/0375-9474(71)90308-3} {\bibfield
  {journal} {\bibinfo  {journal} {Nucl. Phys. A}\ }\textbf {\bibinfo {volume}
  {177}},\ \bibinfo {pages} {559} (\bibinfo {year} {1971})}\BibitemShut
  {NoStop}%
\bibitem [{\citenamefont {Dover}\ and\ \citenamefont {Giai}(1972)}]{dov72}%
  \BibitemOpen
  \bibfield  {author} {\bibinfo {author} {\bibfnamefont {C.~B.}\ \bibnamefont
  {Dover}}\ and\ \bibinfo {author} {\bibfnamefont {N.~V.}\ \bibnamefont
  {Giai}},\ }\bibfield  {title} {\bibinfo {title} {{The nucleon-nucleus
  potential in the {H}artree-{F}ock approximation with {S}kyrme's
  interaction}},\ }\href
  {https://doi.org/https://doi.org/10.1016/0375-9474(72)90148-0} {\bibfield
  {journal} {\bibinfo  {journal} {Nucl. Phys. A}\ }\textbf {\bibinfo {volume}
  {190}},\ \bibinfo {pages} {373} (\bibinfo {year} {1972})}\BibitemShut
  {NoStop}%
\bibitem [{\citenamefont {Vautherin}\ and\ \citenamefont
  {Brink}(1972)}]{vau72}%
  \BibitemOpen
  \bibfield  {author} {\bibinfo {author} {\bibfnamefont {D.}~\bibnamefont
  {Vautherin}}\ and\ \bibinfo {author} {\bibfnamefont {D.~M.}\ \bibnamefont
  {Brink}},\ }\bibfield  {title} {\bibinfo {title} {{{H}artree-{F}ock
  calculations with {S}kyrme's interaction. {I}. {S}pherical Nuclei}},\ }\href
  {https://doi.org/10.1103/PhysRevC.5.626} {\bibfield  {journal} {\bibinfo
  {journal} {Phys. Rev. C}\ }\textbf {\bibinfo {volume} {5}},\ \bibinfo {pages}
  {626} (\bibinfo {year} {1972})}\BibitemShut {NoStop}%
\bibitem [{\citenamefont {Col\`{o}}\ \emph {et~al.}(2013)\citenamefont
  {Col\`{o}}, \citenamefont {Cao}, \citenamefont {Giai},\ and\ \citenamefont
  {Capelli}}]{col13}%
  \BibitemOpen
  \bibfield  {author} {\bibinfo {author} {\bibfnamefont {G.}~\bibnamefont
  {Col\`{o}}}, \bibinfo {author} {\bibfnamefont {L.}~\bibnamefont {Cao}},
  \bibinfo {author} {\bibfnamefont {N.~V.}\ \bibnamefont {Giai}},\ and\
  \bibinfo {author} {\bibfnamefont {L.}~\bibnamefont {Capelli}},\ }\bibfield
  {title} {\bibinfo {title} {{Self-consistent {RPA} calculations with
  {S}kyrme-type interactions: The {\tt{skyrme\_rpa}} program}},\ }\href
  {https://doi.org/10.1016/j.cpc.2012.07.016} {\bibfield  {journal} {\bibinfo
  {journal} {Comput. Phys. Commun.}\ }\textbf {\bibinfo {volume} {184}},\
  \bibinfo {pages} {142} (\bibinfo {year} {2013})}\BibitemShut {NoStop}%
\bibitem [{\citenamefont {Ajzenberg-Selove}(1991)}]{sel91}%
  \BibitemOpen
  \bibfield  {author} {\bibinfo {author} {\bibfnamefont {F.}~\bibnamefont
  {Ajzenberg-Selove}},\ }\bibfield  {title} {\bibinfo {title} {{Energy levels
  of light nuclei $A$ = 13-15}},\ }\href
  {https://doi.org/https://doi.org/10.1016/0375-9474(91)90446-D} {\bibfield
  {journal} {\bibinfo  {journal} {Nucl. Phys. A}\ }\textbf {\bibinfo {volume}
  {523}},\ \bibinfo {pages} {1} (\bibinfo {year} {1991})}\BibitemShut {NoStop}%
\bibitem [{\citenamefont {Tilley}\ \emph {et~al.}(1993)\citenamefont {Tilley},
  \citenamefont {Weller},\ and\ \citenamefont {Cheves}}]{til93}%
  \BibitemOpen
  \bibfield  {author} {\bibinfo {author} {\bibfnamefont {D.~R.}\ \bibnamefont
  {Tilley}}, \bibinfo {author} {\bibfnamefont {H.~R.}\ \bibnamefont {Weller}},\
  and\ \bibinfo {author} {\bibfnamefont {C.~M.}\ \bibnamefont {Cheves}},\
  }\bibfield  {title} {\bibinfo {title} {{Energy levels of light nuclei $A$ =
  16-17}},\ }\href
  {https://doi.org/https://doi.org/10.1016/0375-9474(93)90073-7} {\bibfield
  {journal} {\bibinfo  {journal} {Nucl. Phys. A}\ }\textbf {\bibinfo {volume}
  {564}},\ \bibinfo {pages} {1} (\bibinfo {year} {1993})}\BibitemShut {NoStop}%
\bibitem [{\citenamefont {Wang}\ \emph {et~al.}(2012)\citenamefont {Wang},
  \citenamefont {Audi}, \citenamefont {Wapstra}, \citenamefont {Kondev},
  \citenamefont {MacCormick}, \citenamefont {Xu},\ and\ \citenamefont
  {Pfeiffer}}]{wan12}%
  \BibitemOpen
  \bibfield  {author} {\bibinfo {author} {\bibfnamefont {M.}~\bibnamefont
  {Wang}}, \bibinfo {author} {\bibfnamefont {G.}~\bibnamefont {Audi}}, \bibinfo
  {author} {\bibfnamefont {A.~H.}\ \bibnamefont {Wapstra}}, \bibinfo {author}
  {\bibfnamefont {F.~G.}\ \bibnamefont {Kondev}}, \bibinfo {author}
  {\bibfnamefont {M.}~\bibnamefont {MacCormick}}, \bibinfo {author}
  {\bibfnamefont {X.}~\bibnamefont {Xu}},\ and\ \bibinfo {author}
  {\bibfnamefont {B.}~\bibnamefont {Pfeiffer}},\ }\bibfield  {title} {\bibinfo
  {title} {{The Ame2012 atomic mass evaluation}},\ }\href
  {https://doi.org/10.1088/1674-1137/36/12/003} {\bibfield  {journal} {\bibinfo
   {journal} {Chin. Phys. C}\ }\textbf {\bibinfo {volume} {36}},\ \bibinfo
  {pages} {1603} (\bibinfo {year} {2012})}\BibitemShut {NoStop}%
\bibitem [{\citenamefont {Macfarlane}\ and\ \citenamefont
  {French}(1960)}]{mac60}%
  \BibitemOpen
  \bibfield  {author} {\bibinfo {author} {\bibfnamefont {M.~H.}\ \bibnamefont
  {Macfarlane}}\ and\ \bibinfo {author} {\bibfnamefont {J.~B.}\ \bibnamefont
  {French}},\ }\bibfield  {title} {\bibinfo {title} {{Stripping Reactions and
  the Structure of Light and Intermediate Nuclei}},\ }\href
  {https://doi.org/10.1103/RevModPhys.32.567} {\bibfield  {journal} {\bibinfo
  {journal} {Rev. Mod. Phys.}\ }\textbf {\bibinfo {volume} {32}},\ \bibinfo
  {pages} {567} (\bibinfo {year} {1960})}\BibitemShut {NoStop}%
\bibitem [{\citenamefont {Cohen}\ and\ \citenamefont {Kurath}(1967)}]{coh67}%
  \BibitemOpen
  \bibfield  {author} {\bibinfo {author} {\bibfnamefont {S.}~\bibnamefont
  {Cohen}}\ and\ \bibinfo {author} {\bibfnamefont {D.}~\bibnamefont {Kurath}},\
  }\bibfield  {title} {\bibinfo {title} {{Spectroscopic factors for the $1p$
  shell}},\ }\href
  {https://doi.org/https://doi.org/10.1016/0375-9474(67)90285-0} {\bibfield
  {journal} {\bibinfo  {journal} {Nucl. Phys. A}\ }\textbf {\bibinfo {volume}
  {101}},\ \bibinfo {pages} {1} (\bibinfo {year} {1967})}\BibitemShut {NoStop}%
\bibitem [{\citenamefont {Hao}\ \emph {et~al.}(2015)\citenamefont {Hao},
  \citenamefont {Loc},\ and\ \citenamefont {Phuc}}]{hao15}%
  \BibitemOpen
  \bibfield  {author} {\bibinfo {author} {\bibfnamefont {T.~V.~N.}\
  \bibnamefont {Hao}}, \bibinfo {author} {\bibfnamefont {B.~M.}\ \bibnamefont
  {Loc}},\ and\ \bibinfo {author} {\bibfnamefont {N.~H.}\ \bibnamefont
  {Phuc}},\ }\bibfield  {title} {\bibinfo {title} {{Low-energy nucleon-nucleus
  scattering within the energy density functional approach}},\ }\href
  {https://doi.org/10.1103/PhysRevC.92.014605} {\bibfield  {journal} {\bibinfo
  {journal} {Phys. Rev. C}\ }\textbf {\bibinfo {volume} {92}},\ \bibinfo
  {pages} {014605} (\bibinfo {year} {2015})}\BibitemShut {NoStop}%
\bibitem [{\citenamefont {Kitazawa}\ \emph {et~al.}(1998)\citenamefont
  {Kitazawa}, \citenamefont {Go},\ and\ \citenamefont {Igashira}}]{kit98}%
  \BibitemOpen
  \bibfield  {author} {\bibinfo {author} {\bibfnamefont {H.}~\bibnamefont
  {Kitazawa}}, \bibinfo {author} {\bibfnamefont {K.}~\bibnamefont {Go}},\ and\
  \bibinfo {author} {\bibfnamefont {M.}~\bibnamefont {Igashira}},\ }\bibfield
  {title} {\bibinfo {title} {{Low-energy neutron direct capture by
  ${}^{12}\mathrm{C}$ in a dispersive optical potential}},\ }\href
  {https://doi.org/10.1103/PhysRevC.57.202} {\bibfield  {journal} {\bibinfo
  {journal} {Phys. Rev. C}\ }\textbf {\bibinfo {volume} {57}},\ \bibinfo
  {pages} {202} (\bibinfo {year} {1998})}\BibitemShut {NoStop}%
\bibitem [{\citenamefont {Tian}\ \emph {et~al.}(2018)\citenamefont {Tian},
  \citenamefont {Pang},\ and\ \citenamefont {Ma}}]{tia18}%
  \BibitemOpen
  \bibfield  {author} {\bibinfo {author} {\bibfnamefont {Y.}~\bibnamefont
  {Tian}}, \bibinfo {author} {\bibfnamefont {D.~Y.}\ \bibnamefont {Pang}},\
  and\ \bibinfo {author} {\bibfnamefont {Z.-y.}\ \bibnamefont {Ma}},\
  }\bibfield  {title} {\bibinfo {title} {{Effects of nonlocality of nuclear
  potentials on direct capture reactions}},\ }\href
  {https://doi.org/10.1103/PhysRevC.97.064615} {\bibfield  {journal} {\bibinfo
  {journal} {Phys. Rev. C}\ }\textbf {\bibinfo {volume} {97}},\ \bibinfo
  {pages} {064615} (\bibinfo {year} {2018})}\BibitemShut {NoStop}%
\bibitem [{\citenamefont {Morlock}\ \emph {et~al.}(1997)\citenamefont
  {Morlock}, \citenamefont {Kunz}, \citenamefont {Mayer}, \citenamefont
  {Jaeger}, \citenamefont {M\"uller}, \citenamefont {Hammer}, \citenamefont
  {Mohr}, \citenamefont {Oberhummer}, \citenamefont {Staudt},\ and\
  \citenamefont {K\"olle}}]{mor97}%
  \BibitemOpen
  \bibfield  {author} {\bibinfo {author} {\bibfnamefont {R.}~\bibnamefont
  {Morlock}}, \bibinfo {author} {\bibfnamefont {R.}~\bibnamefont {Kunz}},
  \bibinfo {author} {\bibfnamefont {A.}~\bibnamefont {Mayer}}, \bibinfo
  {author} {\bibfnamefont {M.}~\bibnamefont {Jaeger}}, \bibinfo {author}
  {\bibfnamefont {A.}~\bibnamefont {M\"uller}}, \bibinfo {author}
  {\bibfnamefont {J.~W.}\ \bibnamefont {Hammer}}, \bibinfo {author}
  {\bibfnamefont {P.}~\bibnamefont {Mohr}}, \bibinfo {author} {\bibfnamefont
  {H.}~\bibnamefont {Oberhummer}}, \bibinfo {author} {\bibfnamefont
  {G.}~\bibnamefont {Staudt}},\ and\ \bibinfo {author} {\bibfnamefont
  {V.}~\bibnamefont {K\"olle}},\ }\bibfield  {title} {\bibinfo {title} {{Halo
  Properties of the First $1/{2}^{+}$ State in ${}^{17}\mathrm{F}$ from the
  ${}^{16}\mathrm{O}(p,\gamma){}^{17}\mathrm{F}$ Reaction}},\ }\href
  {https://doi.org/10.1103/PhysRevLett.79.3837} {\bibfield  {journal} {\bibinfo
   {journal} {Phys. Rev. Lett.}\ }\textbf {\bibinfo {volume} {79}},\ \bibinfo
  {pages} {3837} (\bibinfo {year} {1997})}\BibitemShut {NoStop}%
\bibitem [{\citenamefont {Bertulani}\ and\ \citenamefont
  {Danielewicz}(2003)}]{ber03}%
  \BibitemOpen
  \bibfield  {author} {\bibinfo {author} {\bibfnamefont {C.}~\bibnamefont
  {Bertulani}}\ and\ \bibinfo {author} {\bibfnamefont {P.}~\bibnamefont
  {Danielewicz}},\ }\bibfield  {title} {\bibinfo {title} {{Breakup of the
  weakly bound $^{17}\mathrm{F}$ nucleus}},\ }\href
  {https://doi.org/https://doi.org/10.1016/S0375-9474(03)00640-7} {\bibfield
  {journal} {\bibinfo  {journal} {Nucl. Phys. A}\ }\textbf {\bibinfo {volume}
  {717}},\ \bibinfo {pages} {199 } (\bibinfo {year} {2003})}\BibitemShut
  {NoStop}%
\bibitem [{\citenamefont {Igashira}\ \emph {et~al.}(1995)\citenamefont
  {Igashira}, \citenamefont {Nagai}, \citenamefont {Masuda}, \citenamefont
  {Ohsaki},\ and\ \citenamefont {Kitazawa}}]{iga95}%
  \BibitemOpen
  \bibfield  {author} {\bibinfo {author} {\bibfnamefont {M.}~\bibnamefont
  {Igashira}}, \bibinfo {author} {\bibfnamefont {Y.}~\bibnamefont {Nagai}},
  \bibinfo {author} {\bibfnamefont {K.}~\bibnamefont {Masuda}}, \bibinfo
  {author} {\bibfnamefont {T.}~\bibnamefont {Ohsaki}},\ and\ \bibinfo {author}
  {\bibfnamefont {H.}~\bibnamefont {Kitazawa}},\ }\bibfield  {title} {\bibinfo
  {title} {{Measurement of the ${}^{16}\mathrm{O}(n,\gamma){}^{17}\mathrm{O}$
  reaction cross section at stellar energy and the critical role of nonresonant
  $p$-wave neutron capture}},\ }\href {https://doi.org/doi = {10.1086/187797}}
  {\bibfield  {journal} {\bibinfo  {journal} {Astrophys. J.}\ }\textbf
  {\bibinfo {volume} {441}},\ \bibinfo {pages} {L89} (\bibinfo {year}
  {1995})}\BibitemShut {NoStop}%
\bibitem [{\citenamefont {Nakada}\ \emph {et~al.}(1971)\citenamefont {Nakada},
  \citenamefont {Torizuka},\ and\ \citenamefont {Horikawa}}]{nak71}%
  \BibitemOpen
  \bibfield  {author} {\bibinfo {author} {\bibfnamefont {A.}~\bibnamefont
  {Nakada}}, \bibinfo {author} {\bibfnamefont {Y.}~\bibnamefont {Torizuka}},\
  and\ \bibinfo {author} {\bibfnamefont {Y.}~\bibnamefont {Horikawa}},\
  }\bibfield  {title} {\bibinfo {title} {{Determination of the Deformation in
  $^{12}\mathrm{C}$ from Electron Scattering}},\ }\href
  {https://doi.org/10.1103/PhysRevLett.27.745} {\bibfield  {journal} {\bibinfo
  {journal} {Phys. Rev. Lett.}\ }\textbf {\bibinfo {volume} {27}},\ \bibinfo
  {pages} {745} (\bibinfo {year} {1971})}\BibitemShut {NoStop}%
\bibitem [{\citenamefont {Timofeyuk}(2013)}]{tim13}%
  \BibitemOpen
  \bibfield  {author} {\bibinfo {author} {\bibfnamefont {N.~K.}\ \bibnamefont
  {Timofeyuk}},\ }\bibfield  {title} {\bibinfo {title} {{Spectroscopic factors
  and asymptotic normalization coefficients for $0p$-shell nuclei: Recent
  updates}},\ }\href {https://doi.org/10.1103/PhysRevC.88.044315} {\bibfield
  {journal} {\bibinfo  {journal} {Phys. Rev. C}\ }\textbf {\bibinfo {volume}
  {88}},\ \bibinfo {pages} {044315} (\bibinfo {year} {2013})}\BibitemShut
  {NoStop}%
\bibitem [{\citenamefont {Bailey}\ and\ \citenamefont
  {Stratton}(1950)}]{bai50}%
  \BibitemOpen
  \bibfield  {author} {\bibinfo {author} {\bibfnamefont {C.~L.}\ \bibnamefont
  {Bailey}}\ and\ \bibinfo {author} {\bibfnamefont {W.~R.}\ \bibnamefont
  {Stratton}},\ }\bibfield  {title} {\bibinfo {title} {{Cross Section of the
  ${}^{12}\mathrm{C}(p,\ensuremath{\gamma}){}^{13}\mathrm{N}$ Reaction at Low
  Energies}},\ }\href {https://doi.org/10.1103/PhysRev.77.194} {\bibfield
  {journal} {\bibinfo  {journal} {Phys. Rev.}\ }\textbf {\bibinfo {volume}
  {77}},\ \bibinfo {pages} {194} (\bibinfo {year} {1950})}\BibitemShut
  {NoStop}%
\bibitem [{\citenamefont {Burtebaev}\ \emph {et~al.}(2008)\citenamefont
  {Burtebaev}, \citenamefont {Igamov}, \citenamefont {Peterson}, \citenamefont
  {Yarmukhamedov},\ and\ \citenamefont {Zazulin}}]{bur08}%
  \BibitemOpen
  \bibfield  {author} {\bibinfo {author} {\bibfnamefont {N.}~\bibnamefont
  {Burtebaev}}, \bibinfo {author} {\bibfnamefont {S.~B.}\ \bibnamefont
  {Igamov}}, \bibinfo {author} {\bibfnamefont {R.~J.}\ \bibnamefont
  {Peterson}}, \bibinfo {author} {\bibfnamefont {R.}~\bibnamefont
  {Yarmukhamedov}},\ and\ \bibinfo {author} {\bibfnamefont {D.~M.}\
  \bibnamefont {Zazulin}},\ }\bibfield  {title} {\bibinfo {title} {{New
  measurements of the astrophysical ${S}$ factor for
  ${}^{12}\mathrm{C}(p,\ensuremath{\gamma}){}^{13}\mathrm{N}$ reaction at low
  energies and the asymptotic normalization coefficient (nuclear vertex
  constant) for the
  $p+{}^{12}\mathrm{C}\ensuremath{\rightarrow}{}^{13}\mathrm{N}$ reaction}},\
  }\href {https://doi.org/10.1103/PhysRevC.78.035802} {\bibfield  {journal}
  {\bibinfo  {journal} {Phys. Rev. C}\ }\textbf {\bibinfo {volume} {78}},\
  \bibinfo {pages} {035802} (\bibinfo {year} {2008})}\BibitemShut {NoStop}%
\bibitem [{\citenamefont {Rolfs}\ and\ \citenamefont {Azuma}(1974)}]{rol74}%
  \BibitemOpen
  \bibfield  {author} {\bibinfo {author} {\bibfnamefont {C.}~\bibnamefont
  {Rolfs}}\ and\ \bibinfo {author} {\bibfnamefont {R.~E.}\ \bibnamefont
  {Azuma}},\ }\bibfield  {title} {\bibinfo {title} {{Interference effects in
  ${}^{12}\mathrm{C}(p,\gamma){}^{13}\mathrm{N}$ and direct capture to unbound
  states}},\ }\href
  {https://doi.org/https://doi.org/10.1016/0375-9474(74)90798-2} {\bibfield
  {journal} {\bibinfo  {journal} {Nucl. Phys. A}\ }\textbf {\bibinfo {volume}
  {227}},\ \bibinfo {pages} {291} (\bibinfo {year} {1974})}\BibitemShut
  {NoStop}%
\bibitem [{\citenamefont {Vogl}(1963)}]{vog63}%
  \BibitemOpen
  \bibfield  {author} {\bibinfo {author} {\bibfnamefont {J.~L.}\ \bibnamefont
  {Vogl}},\ }\emph {\bibinfo {title} {{Radiative capture of protons by
  $\mathrm{C}^{12}$ and $\mathrm{C}^{13}$ below 700 keV}}},\ \href@noop {}
  {Ph.D. thesis},\ \bibinfo  {school} {California Institute of Technology}
  (\bibinfo {year} {1963})\BibitemShut {NoStop}%
\bibitem [{\citenamefont {Kikuchi}\ \emph {et~al.}(1998)\citenamefont
  {Kikuchi}, \citenamefont {Nagai}, \citenamefont {Suzuki}, \citenamefont
  {Shima}, \citenamefont {Kii}, \citenamefont {Igashira}, \citenamefont
  {Mengoni},\ and\ \citenamefont {Otsuka}}]{kik98}%
  \BibitemOpen
  \bibfield  {author} {\bibinfo {author} {\bibfnamefont {T.}~\bibnamefont
  {Kikuchi}}, \bibinfo {author} {\bibfnamefont {Y.}~\bibnamefont {Nagai}},
  \bibinfo {author} {\bibfnamefont {T.~S.}\ \bibnamefont {Suzuki}}, \bibinfo
  {author} {\bibfnamefont {T.}~\bibnamefont {Shima}}, \bibinfo {author}
  {\bibfnamefont {T.}~\bibnamefont {Kii}}, \bibinfo {author} {\bibfnamefont
  {M.}~\bibnamefont {Igashira}}, \bibinfo {author} {\bibfnamefont
  {A.}~\bibnamefont {Mengoni}},\ and\ \bibinfo {author} {\bibfnamefont
  {T.}~\bibnamefont {Otsuka}},\ }\bibfield  {title} {\bibinfo {title}
  {{Nonresonant direct $p$- and $d$-wave neutron capture by
  ${}^{12}\mathrm{C}$}},\ }\href {https://doi.org/10.1103/PhysRevC.57.2724}
  {\bibfield  {journal} {\bibinfo  {journal} {Phys. Rev. C}\ }\textbf {\bibinfo
  {volume} {57}},\ \bibinfo {pages} {2724} (\bibinfo {year}
  {1998})}\BibitemShut {NoStop}%
\bibitem [{\citenamefont {{Ohsaki}}\ \emph {et~al.}(1994)\citenamefont
  {{Ohsaki}}, \citenamefont {{Nagai}}, \citenamefont {{Igashira}},
  \citenamefont {{Shima}}, \citenamefont {{Takeda}}, \citenamefont {{Seino}},\
  and\ \citenamefont {{Irie}}}]{ohs94}%
  \BibitemOpen
  \bibfield  {author} {\bibinfo {author} {\bibfnamefont {T.}~\bibnamefont
  {{Ohsaki}}}, \bibinfo {author} {\bibfnamefont {Y.}~\bibnamefont {{Nagai}}},
  \bibinfo {author} {\bibfnamefont {M.}~\bibnamefont {{Igashira}}}, \bibinfo
  {author} {\bibfnamefont {T.}~\bibnamefont {{Shima}}}, \bibinfo {author}
  {\bibfnamefont {K.}~\bibnamefont {{Takeda}}}, \bibinfo {author}
  {\bibfnamefont {S.}~\bibnamefont {{Seino}}},\ and\ \bibinfo {author}
  {\bibfnamefont {T.}~\bibnamefont {{Irie}}},\ }\bibfield  {title} {\bibinfo
  {title} {{New Measurement of the $^{12}\mathrm{C}(n,\gamma)^{13}\mathrm{C}$
  Reaction Cross Section}},\ }\href {https://doi.org/10.1086/173783} {\bibfield
   {journal} {\bibinfo  {journal} {Astrophys. J.}\ }\textbf {\bibinfo {volume}
  {422}},\ \bibinfo {pages} {912} (\bibinfo {year} {1994})}\BibitemShut
  {NoStop}%
\bibitem [{\citenamefont {Otsuka}\ \emph {et~al.}(1994)\citenamefont {Otsuka},
  \citenamefont {Ishihara}, \citenamefont {Fukunishi}, \citenamefont
  {Nakamura},\ and\ \citenamefont {Yokoyama}}]{ots94}%
  \BibitemOpen
  \bibfield  {author} {\bibinfo {author} {\bibfnamefont {T.}~\bibnamefont
  {Otsuka}}, \bibinfo {author} {\bibfnamefont {M.}~\bibnamefont {Ishihara}},
  \bibinfo {author} {\bibfnamefont {N.}~\bibnamefont {Fukunishi}}, \bibinfo
  {author} {\bibfnamefont {T.}~\bibnamefont {Nakamura}},\ and\ \bibinfo
  {author} {\bibfnamefont {M.}~\bibnamefont {Yokoyama}},\ }\bibfield  {title}
  {\bibinfo {title} {{Neutron halo effect on direct neutron capture and
  photodisintegration}},\ }\href {https://doi.org/10.1103/PhysRevC.49.R2289}
  {\bibfield  {journal} {\bibinfo  {journal} {Phys. Rev. C}\ }\textbf {\bibinfo
  {volume} {49}},\ \bibinfo {pages} {R2289} (\bibinfo {year}
  {1994})}\BibitemShut {NoStop}%
\bibitem [{\citenamefont {Takai}\ \emph {et~al.}(1977)\citenamefont {Takai},
  \citenamefont {Kambara}, \citenamefont {Tada}, \citenamefont {Nakamura},\
  and\ \citenamefont {Kobayashi}}]{tak77}%
  \BibitemOpen
  \bibfield  {author} {\bibinfo {author} {\bibfnamefont {M.}~\bibnamefont
  {Takai}}, \bibinfo {author} {\bibfnamefont {T.}~\bibnamefont {Kambara}},
  \bibinfo {author} {\bibfnamefont {K.}~\bibnamefont {Tada}}, \bibinfo {author}
  {\bibfnamefont {M.}~\bibnamefont {Nakamura}},\ and\ \bibinfo {author}
  {\bibfnamefont {S.}~\bibnamefont {Kobayashi}},\ }\bibfield  {title} {\bibinfo
  {title} {{Two-Step Process in the $^{12}\mathrm{C}(d,p)^{13}\mathrm{C}$
  Reaction}},\ }\href {https://doi.org/10.1143/JPSJ.43.17} {\bibfield
  {journal} {\bibinfo  {journal} {J. Phys. Soc. Japan}\ }\textbf {\bibinfo
  {volume} {43}},\ \bibinfo {pages} {17} (\bibinfo {year} {1977})}\BibitemShut
  {NoStop}%
\bibitem [{\citenamefont {Ohnuma}\ \emph {et~al.}(1986)\citenamefont {Ohnuma},
  \citenamefont {Hoshino}, \citenamefont {Mikoshiba}, \citenamefont {Raywood},
  \citenamefont {Sakaguchi}, \citenamefont {Shute}, \citenamefont {Spicer},
  \citenamefont {Tanaka}, \citenamefont {Tanifuji}, \citenamefont {Terasawa},\
  and\ \citenamefont {Yasue}}]{ohn86}%
  \BibitemOpen
  \bibfield  {author} {\bibinfo {author} {\bibfnamefont {H.}~\bibnamefont
  {Ohnuma}}, \bibinfo {author} {\bibfnamefont {N.}~\bibnamefont {Hoshino}},
  \bibinfo {author} {\bibfnamefont {O.}~\bibnamefont {Mikoshiba}}, \bibinfo
  {author} {\bibfnamefont {K.}~\bibnamefont {Raywood}}, \bibinfo {author}
  {\bibfnamefont {A.}~\bibnamefont {Sakaguchi}}, \bibinfo {author}
  {\bibfnamefont {G.~G.}\ \bibnamefont {Shute}}, \bibinfo {author}
  {\bibfnamefont {B.~M.}\ \bibnamefont {Spicer}}, \bibinfo {author}
  {\bibfnamefont {M.~H.}\ \bibnamefont {Tanaka}}, \bibinfo {author}
  {\bibfnamefont {M.}~\bibnamefont {Tanifuji}}, \bibinfo {author}
  {\bibfnamefont {T.}~\bibnamefont {Terasawa}},\ and\ \bibinfo {author}
  {\bibfnamefont {M.}~\bibnamefont {Yasue}},\ }\bibfield  {title} {\bibinfo
  {title} {{The ${}^{12}\mathrm{C}(d, p){}^{13}\mathrm{C}$ reaction at $E_d =
  30$ MeV and positive-parity states in ${}^{13}\mathrm{C}$}},\ }\href
  {https://doi.org/https://doi.org/10.1016/0375-9474(86)90087-4} {\bibfield
  {journal} {\bibinfo  {journal} {Nucl. Phys. A}\ }\textbf {\bibinfo {volume}
  {448}},\ \bibinfo {pages} {205} (\bibinfo {year} {1986})}\BibitemShut
  {NoStop}%
\end{thebibliography}%
\end{document}